\documentclass[a4paper,12pt]{article}
\usepackage[english]{babel}
\usepackage{jheppub2}
\usepackage{subfig}
\usepackage[T1]{fontenc} 
\usepackage{graphicx}
\usepackage{epsfig}
\usepackage{amsmath}
\usepackage{amssymb}
\usepackage{float}
\usepackage{placeins}
\usepackage{braket}
\usepackage{slashed}
\usepackage{mathdots}
\usepackage{lipsum}

\allowdisplaybreaks[4]

\title{Operator mixing in massless QCD-like theories and Poincar\`e-Dulac theorem}

\author[a]{Matteo Becchetti}
\author[b]{Marco Bochicchio}
\affiliation[a]{Physics Department, Torino University and INFN Torino, \\
Via Pietro Giuria 1, I-10125 Torino, Italy}
\affiliation[b]{Physics Department, INFN Roma1, \\
Piazzale A. Moro 2, Roma, I-00185, Italy}
\emailAdd{matteo.becchetti@unito.it}
\emailAdd{marco.bochicchio@roma1.infn.it}

\abstract{Recently, a differential-geometric approach to operator mixing in massless QCD-like theories -- that involves canonical forms, obtained by means of gauge transformations, based on the Poincar\`e-Dulac theorem for the linear system that defines the renormalized mixing matrix in the coordinate representation $Z(x,\mu)$ -- has been proposed in \cite{MB0}. Specifically, it has been determined under which conditions a renormalization scheme exists where the linear system -- and correspondingly $Z(x,\mu)$ -- may be set in a diagonal canonical form that is one-loop exact to all perturbative orders -- the nonresonant diagonalizable $\frac{\gamma_0}{\beta_0}$ case (I) -- according to the Poincar\`e-Dulac theorem. Moreover, the remaining cases, (II), (III) and (IV), of operator mixing, where such diagonalization is not possible, have also been classified in \cite{MB0}. Accordingly, if the matrix $\frac{\gamma_0}{\beta_0}$, with $\gamma(g)=\gamma_0 g^2+\cdots$ the matrix of the anomalous dimensions and $\beta(g)=-\beta_0 g^3 + \cdots$ the beta function, either is diagonalizable but a resonant condition for its eigenvalues and the system holds (II) or is nondiagonalizable and nonresonant (III), or is both nondiagonalizable and resonant (IV), $Z(x,\mu)$ is nondiagonalizable. Yet, we argue that in the gauge-invariant Hermitian sector of a massless QCD-like theory, which should be unitary in its free conformal limit at $g(\mu)=0$, $\frac{\gamma_0}{\beta_0}$ should be diagonalizable, because otherwise, to the order of $g^2(\mu)$, a logCFT would arise that is nonunitary at $g(\mu)=0$. Nevertheless, even if $\frac{\gamma_0}{\beta_0}$ is diagonalizable, the associated linear system may be resonant, thus realizing the alternative (II) above.
In the cases (II), (III) and (IV), where $Z(x,\mu)$ is nondiagonalizable, we demonstrate that its canonical form may be factorized into the exponential of a linear combination of upper triangular nilpotent constant matrices with coefficients that asymptotically in the UV are powers of logs of the running coupling, i.e., powers of loglogs of the coordinates, and a diagonal matrix as in the nonresonant diagonalizable case (I). Hence, its ultraviolet asymptotics differs intrinsically from the case (I) and, for asymptotically free theories, this is the closest analog of logCFTs. We also work out in detail physical realizations of the cases (I) and (II).}

\DeclareMathOperator{\Tr}{Tr}
\newcommand{\beq}{\begin{equation}}
\newcommand{\be}{\begin{equation}
\newcommand{\ee}{\end{equation}}}
\newcommand{\eeq}{\end{equation}}
\newcommand{\nn}{\nonumber}
\newcommand{\bea}{\begin{eqnarray}}
\newcommand{\eea}{\end{eqnarray}}
\newcommand{\bfig}{\begin{figure}}
\newcommand{\efig}{\end{figure}}
\newcommand{\bc}{\begin{center}}
\newcommand{\ec}{\end{center}}

\newcommand{\f}[2]{\frac{#1}{#2}}

\date{}
\begin{document}
\maketitle
\flushbottom

\section{Introduction and physics motivations}

The aim of the present paper is to reconsider the operator mixing and the associated ultraviolet (UV) asymptotics of the renormalized mixing matrix $Z(x,\mu)$ in the coordinate representation in asymptotically free Yang-Mills (YM) theories \footnote{We only consider YM theories with a single gauge coupling. Our methods may extend to theories with multiple couplings, at the price of increasing  mathematical complication.} massless to all perturbative orders (massless QCD-like theories for short), in order to analyze further implications of the differential-geometric approach to operator mixing initiated in \cite{MB0}, where an essential role is played by the Poincar\`e-Dulac theorem \cite{PD1} in the framework of canonical forms \cite{PD0} for linear systems of  differential equations. \par
In fact, $Z(x,\mu)$ is a pivotal ingredient to work out the UV asymptotics of gauge-invariant correlators and OPE coefficients that will be considered in a forthcoming paper \cite{BB2}. \par
One problem addressed in \cite{MB0}, which is hardly discussed in the literature, has been to determine under which conditions the operator mixing may be essentially reduced to the multiplicatively renormalizable case. This is the case (I) -- worked out extensively in \cite{MB0} -- of the classification based on the Poincar\`e-Dulac theorem introduced in \cite{MB0}. \par The remaining cases, (II),(III) and (IV), of the aforementioned classification, where such a reduction is not actually possible, are studied in greater detail in the present paper.\par
There are several physics motivations for doing so, since the UV asymptotics of operator mixing enters a number of applications of the renormalization group (RG), which range from the deep inelastic scattering \cite{Buras1} in QCD to 
the evalutation of the ratio $\frac{\epsilon'}{\epsilon}$ \cite{Buras2, Buras3, Martinelli} for the possible implications of new physics -- if any -- and to the constraints \cite{MBM,MBN,MBH,MBR,MBL,BB} to the eventual nonperturbative solution of the large-$N$ limit  \cite{H,V,Migdal,W} of massless QCD-like theories.\par
A further motivation for working out such asymptotics, for the general case of operator mixing as well, occurs as a part of the program, christened the asymptotically free bootstrap in \cite{MBH}, of verifying whether a candidate \cite{MBH} nonperturbative S matrix arises from the nonperturbative 't Hooft large-$N$ expansion \cite{H} of a massless QCD-like theory, where the multiplicative renormalization of gauge-invariant operators has already played a key role \cite{MBN}. \par
In view of the physics applications of operator mixing mentioned above, the key step in \cite{MB0} to obtain the aforementioned UV asymptotics of $Z(x,\mu)$ has been the choice of a suitable renormalization scheme. \par
In this respect, it has been known for some time that exploiting the freedom of changing renormalization scheme may lead to significant advantages.\par
Perhaps, the most famous example is the 't Hooft scheme \cite{HH}, where all the coefficients of the beta function, $\beta(g) = - \beta_0 g^3 - \beta_1 g^5 + \cdots$, but the first two, $\beta_0, \beta_1$, may be set to $0$ by a suitable (formal \footnote{A formal series is not assumed to be convergent. In QCD-like theories, this is appropriate in perturbation theory, which is believed to be only asymptotic in the UV thanks to the asymptotic freedom.}) holomorphic reparametrization of the gauge coupling.\par
In fact, the aforementioned coefficients, $\beta_2, \beta_3, \cdots $, may be set to an arbitrary value by a reparametrization of the coupling,
and this freedom has been exploited in various contexts \cite{Kataev1,Kataev2,Kataev3,Kataev4,Kataev5,Kataev6}, including the supersymmetric one in relation to the exact NSVZ beta function \cite{NSVZ}. \par
Another example is the possibility to set to $0$ all the coefficients but the first one, $\gamma_0$, of the anomalous dimension, $\gamma(g) = \gamma_0 g^2+\cdots$, of a multiplicatively renormalizable operator by a similar \cite{Collins} -- but in general different -- reparametrization of the coupling. \par
These examples are well known, but are not relevant in the present paper:
They exploit the freedom of making (formal) holomorphic diffeomorphisms in the space of the coupling, while the change of scheme that we refer to is actually the (formal) holomorphic nonabelian \footnote{The aforementioned gauge freedom is nonabelian for the mixing of $2$ or more operators, as opposed to the Abelian gauge symmetry in the case of a multiplicatively renormalizable operator. Its nonabelian and (formal) holomorphic character implies the peculiar features described in the present paper.} gauge freedom \cite{MB0} in the choice of the basis of operators that mix under renormalization, in a way that we summarize as follows.

\section{A summary of \cite{MB0}} \label{2}

The key idea in \cite{MB0} has been to employ the time-honored theory of canonical forms \cite{PD0} -- obtained by (formal) holomorphic gauge transformations -- for linear systems of differential equations -- specifically, the Poincar\`e-Dulac theorem \cite{PD1} -- in order to find a sufficient condition by which
a renormalization scheme exists where the matrix $-\frac{\gamma(g)}{\beta(g)}$ in eq. \eqref{sys} can be set in the canonical form:
\bea \label{1.60}
-\frac{\gamma(g)}{\beta(g)}=\frac{\gamma_0}{\beta_0}
\frac{1}{g}
\eea
 that is one-loop exact to all orders of perturbation theory, with:
 \begin{equation}
\label{1.6}
\gamma(g)=- \frac{\partial Z}{\partial \log \mu} Z^{-1}=  \gamma_0 g^2  + \gamma_{1} g^4 + \gamma_2 g^6 + \cdots
\end{equation}
the matrix of the anomalous dimensions, and:
\bea \label{200}
\frac{\partial g}{\partial \log \mu}=\beta(g)= -\beta_0 g^3 - \beta_1 g^5 - \beta_2 g^7 + \cdots 
\eea
the beta function, with $g=g(\mu)$ the renormalized coupling.\par
A sufficient condition \cite{MB0} for a renormalization scheme to exist where $-\frac{\gamma(g)}{\beta(g)}$ admits the canonical form in eq. \eqref{1.60} is that 
the eigenvalues $\lambda_1, \lambda_2, \cdots $ of the matrix $\frac{\gamma_0}{\beta_0}$, in nonincreasing order $\lambda_1 \geq \lambda_2 \geq \cdots$, do not differ by a positive even integer:
\bea \label{1.61}
\lambda_i -\lambda_j -2k  \neq 0
\eea 
for $i\leq j$ and $k$ a positive integer.\par 
If such a renormalization scheme exists, the mixing has been dubbed nonresonant in \cite{MB0}. Otherwise, it has been dubbed resonant. This terminology in \cite{MB0} derives directly from the application of the Poincar\`e-Dulac theorem to the operator mixing, as we recall in the present paper. \par
Moreover, if in addition $\frac{\gamma_0}{\beta_0}$ is diagonalizable by a further change of the operator basis, the renormalized mixing matrix
in the coordinate representation:
\bea \label{Z}
Z(x, \mu)=P\exp\left(-\int^{g(\mu)}_{g(x)}\frac{\gamma(g)}{\beta(g)}dg\right)
\eea
that enters the solution:
\begin{equation}
\label{2.200}
G(x) =  
Z(x, \mu)\mathcal{G}(x,g(\mu),\mu)Z^T(x, \mu)
\end{equation}
of the Callan-Symanzik equation \cite{MB0,C,S,Pes,Zub,BB2}:
\begin{equation}
\label{2.1}
\left(x \cdot \frac{\partial}{\partial x}+\beta(g)\frac{\partial}{\partial g}+2D\right)G+\gamma(g) \, G+G \, \gamma^T(g)=0
\end{equation}
for $2$-point correlators \footnote{Factors of $Z(x, \mu)$ also enter the solution of the Callan-Symanzik equation for the OPE coefficients \cite{BB2}.} in Euclidean space-time: 
\bea
G_{ik}(x) = \langle O_i(x) O_k(0) \rangle 
\eea
of renormalized local gauge-invariant operators $O_i(x)$ \footnote{We employ the convention that the sum over repeated indices is understood.}:
\bea
O_i= Z_{ik} O_{Bk}
\eea
with $O_{Bk}$ the bare operators that mix \footnote{In fact, gauge-invariant operators also mix with BRST-exact operators and with operators that vanish by the equations of motion (EQM) \cite{M01,M02,M03}. But correlators of gauge-invariant operators with BRST-exact operators vanish, while correlators with EQM operators reduce to contact terms. Therefore, for our purposes it suffices to take into account the mixing of gauge-invariant operators only.} under renormalization and $Z$ the bare mixing matrix,
is diagonalizable as well, and its UV asymptotics reduces in the diagonal basis to the multiplicatively renormalizable case:
\begin{equation}
\label{01.900}
Z_{i}(x, \mu)=\exp\left(\int^{g(\mu)}_{g(x)}\frac{\gamma_{0i}}{\beta_0 g}dg\right) = \left(\frac{g(\mu)}{g(x)}\right)^{\frac{\gamma_{0i}}{\beta_0}}
\end{equation}
with $Z_i(x,\mu)$ and $\gamma_{0i}$ the eigenvalues of the corresponding matrices.\par
The key step in \cite{MB0} to obtain the above result has been the differential-geometric interpretation \cite{MB0} 
of a change of basis of renormalized operators, i.e., of a (finite) change of renormalization scheme:
\bea \label{b}
O'_i(x)=S_{ik}(g) O_k(x)
\eea
as a (formal) holomorphic invertible gauge transformation $S(g)$,
of $A(g)$:
\bea \label{A}
A(g)=-\frac{\gamma(g)}{\beta(g)}= \frac{\gamma_0}{\beta_0} \frac{1}{g}+\cdots
\eea
as a (formal) meromorphic connection, with a Fuchsian singularity \footnote{If a meromorphic connection, $A(g)$, has a Fuchsian singularity at a point, a solution of the corresponding linear system in eq. \eqref{1.700} is regular singular \cite{PD1}, i.e., it satisfies a moderate-grow condition in a neighborhood of the point.} -- i.e., a simple pole -- at $g=0$,
that transforms by the gauge transformation $S(g)$ as:
\bea
A'(g)= S(g)A(g)S^{-1}(g)+ \frac{\partial S(g)}{\partial g} S^{-1}(g)
\eea
of $\mathcal{D}$ as the corresponding covariant derivative:
\bea \label{1.710}
\mathcal{D} =  \frac{\partial}{\partial g} - A(g)
\eea
that defines the linear system:
\bea \label{1.700}
\mathcal{D} X=  \left(\frac{\partial}{\partial g} - A(g)\right) X =0
\eea
whose solution $X(g)$, with a suitable initial condition, is $Z(x, \mu)$, and finally of $Z(x, \mu)$:
\bea
Z(x, \mu)=P\exp\left(\int ^{g(\mu)}_{g(x)} A(g) \, dg\right)
=P\exp\left(-\int^{g(\mu)}_{g(x)}\frac{\gamma(g)}{\beta(g)}dg\right)
\eea
as a Wilson line that transforms as:
\bea
Z'(x, \mu)= S(g(\mu)) Z(x, \mu) S^{-1}(g(x))
\eea
for the gauge transformation $S(g)$. \par
Following the interpretation above, the easiest way to compute the UV asymptotics of $Z(x, \mu)$ consists in setting the meromorphic connection $A(g)$
in a canonical form by a suitable holomorphic gauge transformation according to the Poincar\'e-Dulac theorem.\par
Consequently, the classification in \cite{MB0} of operator mixing is as follows: \par
If a renormalization scheme exists where $ -\frac{\gamma(g)}{\beta(g)}$ can be set in the canonical form of eq. \eqref{1.60}, we refer to the mixing as nonresonant, that by eq. \eqref{1.61} is the generic case.
Otherwise, we refer to the mixing as resonant.\par
Besides, $\frac{\gamma_0}{\beta_0}$ may be either diagonalizable or nondiagonalizable.\par
Therefore, there are four cases of operator mixing:\par
(I) Nonresonant diagonalizable $\frac{\gamma_0}{\beta_0}$. \par
(II) Resonant diagonalizable $\frac{\gamma_0}{\beta_0}$. \par
(III) Nonresonant nondiagonalizable $\frac{\gamma_0}{\beta_0}$. \par
(IV) Resonant nondiagonalizable $\frac{\gamma_0}{\beta_0}$. \par
In the case (I), $Z(x,\mu)$ is diagonalizable \cite{MB0} to all orders of perturbation theory, since the mixing is nonresonant and $\frac{\gamma_0}{\beta_0}$ is diagonalizable.\par
The remaining cases, where $Z(x,\mu)$ is not actually diagonalizable, are analyzed in greater detail in the present paper. \par
We believe, as already remarked in \cite{MB0}, that the geometric interpretation above and the employment of the Poincar\`e-Dulac theorem make the subject of operator mixing in the physics literature more transparent than in previous treatments \cite{Buras1,Buras2,Sonoda,SM}.\par

\section{Plan of the paper}

In section \ref{5}, just as a preamble, we work out by elementary methods three examples of operator mixing that, in the special case of two operators, are paradigmatic of the general case. \par
In section \ref{4}, which contains our main arguments and results, we analyze the four cases, (I), (II), (III) and (IV), in the classification above based on the Poincar\`e-Dulac theorem.
Specifically, we work out the corresponding canonical forms for $-\frac{\gamma(g)}{\beta(g)}$ and $Z(x,\mu)$, and the UV asymptotics of $Z(x,\mu)$. 
Moreover, we argue that the cases (III) and (IV) -- where $\frac{\gamma_0}{\beta_0}$ is nondiagonalizable -- are ruled out by unitarity
of the free conformal limit at $g=0$ in the gauge-invariant Hermitian sector of a massless QCD-like theory.\par
In section \ref{6}, we revisit our elementary computation in section \ref{5} in the light of the Poincar\`e-Dulac theorem, obviously finding perfect agreement.
\par
In section \ref{7}, as an application of the general theory in the present paper, we provide physical realizations of the cases (I) and (II).
Specifically, we demonstrate that the case (II) is actually realized for the mixing of four-quark operators in SU($N$) massless QCD with $N_f=N$ flavors of quarks for every $N \geq 4$. Besides, we show that unitarity is implemented at $g=0$ for the mixing of dimension $8$ operators in large-$N$ SU($N$) YM theory, despite the aforementioned mixing would be potentially nonunitary, since $\frac{\gamma_0}{\beta_0}$ is potentially nondiagonalizable in this case, due to the degeneration of some of its eigenvalues. \par
In appendix \ref{A} we discuss asymptotic versus exact correlators.

\section{A preamble: three examples for the mixing of two operators by elementary methods} \label{5}

In case $A(g)=-\dfrac{\gamma(g)}{\beta(g)}$ is upper triangular, we may compute $Z(x, \mu)$ from:
\beq
\label{5.9}
\dfrac{\partial Z}{\partial g} = A(g) Z
\eeq
directly, avoiding the intricacies of the Poincar\`e-Dulac theorem. \par
Indeed, in this case $A(g)$ may be decomposed into the sum of the diagonal, $A_{\Lambda}(g)$, and nilpotent, $A_N(g)$, contributions:
\bea
 A(g) = A_{\Lambda}(g) + A_N(g)
 \eea
Then, $Z(x,\mu)$ may be computed in the form:
\bea
Z(x,\mu)=Z_{\Lambda} (x,\mu)Z_N(x,\mu)
\eea
provided that:
\beq
\label{5.11}
\dfrac{\partial Z_N}{ \partial g} = Z_{\Lambda}^{-1} A_N(g) Z_{\Lambda}Z_N
\eeq
and: 
\beq
\label{5.9}
\dfrac{\partial Z_{\Lambda}}{ \partial g} = A_{\Lambda}(g) Z_{\Lambda}
\eeq
Therefore:
\beq
\label{5.10}
Z_{\Lambda}(x, \mu) =  \exp \left(\int_{g(x)}^{g(\mu)} A_{\Lambda}(g) dg \right)
\eeq
and:
\beq
\label{5.12}
Z_N(x, \mu)  = P \exp \left(\int^{g(\mu)}_{g(x)} Z_{\Lambda}^{-1} A_N(g) Z_{\Lambda} dg \right)
\eeq
Hence, since $Z_{\Lambda}^{-1}A_N(g)Z_{\Lambda}$ is nilpotent as well, the expansion of the path-ordered exponential for $Z_N(x, \mu)$ terminates 
at a finite order, and $Z(x, \mu)$ is computable in a closed form.

\subsection{Nonresonant versus resonant mixing}

In a massless QCD-like theory, according to eq. \eqref{sys}:
\bea
A(g)=\frac{1}{g} \left(A_0 + \sum^{\infty}_ {n=1} A_{2n} g^{2n} \right)
\eea
with:
\bea \label{AA1}
A_0=\frac{\gamma_0}{\beta_0}
\eea
We work out for two operators three examples \footnote{These examples do not necessarily arise from a massless QCD-like theory. Physical examples are worked out in section \ref{7}.} of mixing, which are paradigmatic of the general case. \par
Firstly, we set:
\be
A_0=\Lambda=
\left(\begin{array}{cc}
\lambda_1 &  0  \\
0 &  \lambda_2
\end{array}\right)
\ee
diagonal, with $\lambda_1 \geq \lambda_2$, i.e., we display the eigenvalues of $A_0$ in nonincreasing order. Clearly, by eq. \eqref{AA1}, this is the case that $\gamma_0$
is diagonalizable.\par
Besides, we set:
\be
A_{2k}=N_{2k}=
\left(\begin{array}{cc}
0 &  \nu_{12}  \\
0 & 0
\end{array}\right)
\ee
upper triangular, with $\nu_{12}$ a real nonvanishing number, and $A_{2n}=0$ for $n \neq k$.
Hence, in our example:
\bea \label{5.13}
\dfrac{\partial Z}{\partial g} = \left(\Lambda g^{-1} + N_{2k} g^{2k-1}\right)  Z
\eea
For $A_0$ diagonal, we consider the following two cases that we dub respectively, according to the terminology in section \ref{2}, nonresonant diagonaliazable $\frac{\gamma_0}{\beta_0}$:  
\bea
\lambda_{1}-\lambda_{2}  \neq 2k
\eea
and resonant diagonalizable $\frac{\gamma_0}{\beta_0}$: 
\bea \label{resex2}
\lambda_{1}-\lambda_{2}= 2k
\eea
with $k$ a positive integer.\par
Secondly, we consider the case:
\bea
\lambda_{1}=\lambda_{2}=\lambda
\eea
and:
\be
A_0=\Lambda+N_0=
\left(\begin{array}{cc}
\lambda &  \nu_{12}  \\
0 &  \lambda
\end{array}\right)
\ee
which we dub, according to the terminology in section \ref{2}, nonresonant nondiagonalizable $\frac{\gamma_0}{\beta_0}$, as now $A_0$ is not diagonalizable, since its eigenvalues coincide and $\nu_{12}$ is assumed to be nonzero.\par
We compute the corresponding $Z(x,\mu)$ by the formulas above.

\subsubsection{Nonresonant diagonalizable $\frac{\gamma_0}{\beta_0}$}

$Z_{\Lambda}$ is diagonal and computed by exploiting eq. \eqref{5.10}:
\bea \label{Zdiag}
\label{ZD}
Z_{\Lambda}(x,\mu) = \exp \int_{g(x)}^{g(\mu)} {\Lambda}\dfrac{dg}{g} = \begin{scriptsize}
\left(\begin{array}{cc}
 \left(\dfrac{g(\mu)}{g(x)}\right)^{\lambda_1} & 0 \\
 0 &  \left(\dfrac{g(\mu)}{g(x)}\right)^{\lambda_2} \\
\end{array}
\right)
\end{scriptsize}
\eea
Moreover, by direct evaluation:
\be
Z_{\Lambda}^{-1} A_N(g) Z_{\Lambda} = \left(\begin{array}{cc}
0 & 
 \nu_{12}g^{2k - 1 -\lambda_1 +\lambda_2}(\mu) g^{\lambda_1-\lambda_2}(x)  
\\
0 & 0
\end{array}\right)
\ee
As a consequence:
\bea
Z_N(x,\mu)&=& P \exp \left(\int_{g(x)}^{g(\mu)} Z_{\Lambda}^{-1} A_N(g) Z_{\Lambda} dg \right) \nonumber \\
&=& \left(\begin{array}{cc}
1 &  
\dfrac{\nu_{12}g^{2k}(x)}{\lambda_1 -\lambda_2 -2k}\left(1 - \left(\dfrac{g(\mu)}{g(x)}\right)^{2k +\lambda_2 -\lambda_1}\right)  
 \\
0 & 1
\end{array}\right) 
\eea 
Finally, combining eq. \eqref{ZD} for $Z_{\Lambda}$ and the above result for $Z_N$, we get:
\bea \label{ab}
Z(x,\mu) & = & \begin{scriptsize}
\left(
\begin{array}{cc}
 \left(\dfrac{g(\mu)}{g(x)}\right)^{\lambda_1} & 0 \\
 0 & \left(\dfrac{g(\mu)}{g(x)}\right)^{\lambda_2}
\end{array}
\right) \end{scriptsize}\left(I + \left(\begin{array}{cc}
0 & \dfrac{\nu_{12}g^{2k}(x)}{\lambda_1 -\lambda_2 -2k}\left(1 - \left(\dfrac{g(\mu)}{g(x)}\right)^{2k +\lambda_2 -\lambda_1}\right) \\
0 & 0
\end{array}\right)\right) \nn \\
&=& \begin{scriptsize}\left(
\begin{array}{cc}
 \left(\dfrac{g(\mu)}{g(x)}\right)^{\lambda_1} & \dfrac{\nu_{12}g^{2k}(x)}{\lambda_1-\lambda_2 -2k}\left(\left(\dfrac{g(\mu)}{g(x)}\right)^{\lambda_1} - \left(\dfrac{g(\mu)}{g(x)}\right)^{\lambda_2 + 2k}\right) \\
 0 & \left(\dfrac{g(\mu)}{g(x)}\right)^{\lambda_2}
\end{array}
\right)
\end{scriptsize} \nn \\
&=&\left(\begin{array}{cc}
1 & \dfrac{\nu_{12}g^{2k}(\mu)}{2k-\lambda_1+\lambda_2} \\
0 & 1
\end{array}\right)\begin{scriptsize}\left(\begin{array}{cc}
\left(\dfrac{g(\mu)}{g(x)}\right)^{\lambda_1} & 0 \\
 0 & \left(\dfrac{g(\mu)}{g(x)}\right)^{\lambda_2}
\end{array}\right)\end{scriptsize}\left(\begin{array}{cc}
1 & -\dfrac{\nu_{12}g^{2k}(x)}{2k-\lambda_1+\lambda_2} \\
0 & 1
\end{array}\right) \nn \\
&=&  S(g(\mu))Z_{\Lambda}(x,\mu)S^{-1}(g(x))
\eea
with the holomorphic gauge transformation:
\bea
S(g)= \left(\begin{array}{cc}
1 & \dfrac{\nu_{12}g^{2k}}{2k-\lambda_1+\lambda_2} \\
0 & 1
\end{array}\right)
\eea
Hence, $Z(x, \mu)$ is gauge equivalent to the diagonal $Z_{\Lambda}(x, \mu)$.

\subsubsection{Resonant diagonalizable $\frac{\gamma_0}{\beta_0}$}

$Z_{\Lambda}$ is given again by eq. \eqref{Zdiag},
but now:
\bea
Z_{\Lambda}^{-1} A_N(g) Z_{\Lambda} = \left(\begin{array}{cc}
0 & \nu_{12}\dfrac{g^{2k}(x)}{g(\mu)} \\
0 & 0
\end{array}\right)
\eea
It follows from eq. \eqref{5.12} that:
\bea
Z_N(x,\mu) &=& P \exp \left(\int^{g(\mu)}_{g(x)} Z_{\Lambda}^{-1} A_N(g) Z_{\Lambda} dg \right)\nonumber \\
&=& 
\left(
\begin{array}{cc}
 1 & \nu_{12} g^{2k}(x) \log \dfrac{g(\mu)}{g(x)} \\
 0 & 1 \\
\end{array}
\right)
\eea
Combining the result for $Z_{\Lambda}(x,\mu)$ and $Z_N(x,\mu)$, we obtain:
\bea \label{zsol2}
Z(x,\mu) &=& \begin{scriptsize}
\left(
\begin{array}{cc}
 \left(\dfrac{g(\mu)}{g(x)}\right)^{\lambda_1} & 0 \\
 0 & \left(\dfrac{g(\mu)}{g(x)}\right)^{\lambda_2}
\end{array}
\right) \end{scriptsize} \left(I + \left(\begin{array}{cc}
0 & \nu_{12}g^{2k}(x) \log\dfrac{g(\mu)}{g(x)} \\
0 & 0
\end{array}\right) \right) \nn \\
&=&\begin{scriptsize}\left(
\begin{array}{cc}
 \left(\dfrac{g(\mu)}{g(x)}\right)^{\lambda_1} & \nu_{12}g^{2k}(x) \left(\dfrac{g(\mu)}{g(x)}\right)^{\lambda_1} \log\dfrac{g(\mu)}{g(x)} \\
 0 & \left(\dfrac{g(\mu)}{g(x)}\right)^{\lambda_2} 
\end{array}
\right)
\end{scriptsize} \nn \\
& = &  \left(I + \left(\begin{array}{cc}
0 & \nu_{12}g^{2k}(\mu) \log\dfrac{g(\mu)}{g(x)} \\
0 & 0
\end{array}\right) \right) \begin{scriptsize}
\left(
\begin{array}{cc}
 \left(\dfrac{g(\mu)}{g(x)}\right)^{\lambda_1} & 0 \\
 0 & \left(\dfrac{g(\mu)}{g(x)}\right)^{\lambda_2}
\end{array}
\right) \end{scriptsize}
\eea
Because of the occurrence of the term containing $g^{2k}(\mu) \log \dfrac{g(\mu)}{g(x)}$, $Z(x,\mu)$ is not diagonalizable by a holomorphic gauge transformation. \par

\subsubsection{Nonresonant nondiagonalizable $\frac{\gamma_0}{\beta_0}$}

Finally, we specialize the case above to $k=0$. We get for $\lambda_1=\lambda_2=\lambda$:
\bea
\label{Zsol}
Z(x,\mu) &=& \begin{scriptsize}
\left(
\begin{array}{cc}
 \left(\dfrac{g(\mu)}{g(x)}\right)^{\lambda} & 0 \\
 0 & \left(\dfrac{g(\mu)}{g(x)}\right)^{\lambda}
\end{array}
\right) \end{scriptsize}\left(I + \left(\begin{array}{cc}
0 & \nu_{12} \log \dfrac{g(\mu)}{g(x)} \\
0 & 0
\end{array}\right)\right) \nn \\
&=&\begin{scriptsize}\left(
\begin{array}{cc}
 \left(\dfrac{g(\mu)}{g(x)}\right)^{\lambda} & \nu_{12}\left(\dfrac{g(\mu)}{g(x)}\right)^{\lambda}\log \dfrac{g(\mu)}{g(x)} \\
 0 & \left(\dfrac{g(\mu)}{g(x)}\right)^{\lambda}
\end{array}
\right)
\end{scriptsize} \nn \\
&=& \left(I + \left(\begin{array}{cc}
0 & \nu_{12} \log \dfrac{g(\mu)}{g(x)} \\
0 & 0
\end{array}\right)\right) \begin{scriptsize}
\left(
\begin{array}{cc}
 \left(\dfrac{g(\mu)}{g(x)}\right)^{\lambda} & 0 \\
 0 & \left(\dfrac{g(\mu)}{g(x)}\right)^{\lambda}
\end{array}
\right) \end{scriptsize}
\eea
Again, because of the occurrence of the term containing $\log \dfrac{g(\mu)}{g(x)}$, $Z(x,\mu)$ is not diagonalizable by a holomorphic gauge transformation.

\section{Main arguments and results: operator mixing by the Poincar\`e-Dulac theorem} \label{4}

\subsection{Resonant canonical form of $-\frac{\gamma(g)}{\beta(g)}$ by the Poincar\`e-Dulac theorem} 

In a massless QCD-like theory the meromorphic connection \cite{MB0} $A(g)$ (section \ref{2}) admits the (formal) expansion:
\bea \label{sys}
A(g)=-\frac{\gamma(g)}{\beta(g)}= \frac{1}{g} \left(A_0 + \sum^{\infty}_ {k=1} A_{2k} g^{2k} \right)
\eea
in odd powers of $g$, with the first few coefficients given by:
\bea \label{AA}
A_0&=&\frac{\gamma_0}{\beta_0}
\eea
\bea
A_2&=& \dfrac{\beta_0 \gamma_1 - \beta_1 \gamma_0}{\beta_0^2}
\eea
\bea
 A_4 &=& \dfrac{\beta_1^2 \gamma_0 -\beta_0\beta_2\gamma_0 - \beta_0\beta_1\gamma_1 + \beta_0^2 \gamma_2}{\beta_0^3}
\eea
In general, by the Poincar\`e-Dulac theorem \cite{PD1} (section \ref{PD}), $A(g)$ can be set by a (formal) holomorphic invertible gauge transformation in the canonical resonant form:
\bea \label{resft}
A'(g)=\frac{1}{g} \left(\Lambda+N_0 +  \sum_{k=1} N_{2k} g^{2k} \right)
\eea
where:
\bea \label{J}
A_0=\Lambda+N_0
\eea
is upper triangular, with eigenvalues $\text{diag}(\lambda_1, \lambda_2, \cdots)=\Lambda$ in nonincreasing order $\lambda_1 \geq \lambda_2 \geq \cdots$, and nilpotent part, $N_0$, in normal Jordan form. The upper triangular nilpotent matrices $N_{2k}$ satisfy:
\bea \label{nil}
g^{\Lambda} N_{2k} g^{-\Lambda} = g^{2k} N_{2k}
\eea
i.e., their only nonzero entries, $(N_{2k})_{ij}$, are such that:
\bea \label{ein1}
\lambda_i - \lambda_j=2k
\eea
for $i < j$ and $k$ a positive integer.
Besides, also:
\bea
g^{\Lambda} N_{0} g^{-\Lambda} =  N_{0}
\eea
since $[\Lambda,N_0]=0$ according to the Jordan normal form of $A_0$. \par
Moreover, the sum in eq. \eqref{resft} contains only a finite number of terms, contrary to eq. \eqref{sys}.
Indeed, the number of differences of the eigenvalues is finite, and therefore, because of eq. \eqref{nil}, so it is the number of terms in eq. \eqref{resft}.\par
Eq. \eqref{ein1} is the resonance condition for the eigenvalues of the linear system associated to $A(g)$:
\bea \label{lin}
\mathcal{D} X=  \left(\frac{\partial}{\partial g} - A(g)\right) X =0
\eea
whose solution with a suitable initial condition (sections \ref{F} and \ref{ZZ}) is $Z(x,\mu)$.\par
In fact, from the proof (section \ref{PD}) of the Poincar\`e-Dulac theorem it follows that, once $A_0$ has been set in Jordan normal form by a global 
\footnote{Namely, a constant gauge transformation, i.e., a $g$-independent transformation in our framework.} gauge transformation, precisely only the resonant terms in eq. \eqref{resft} may survive after the gauge transformation that sets eq. \eqref{sys} in the aforementioned canonical form. \par
In this case, the linear system is resonant and consequently the associated operator mixing has been dubbed resonant in \cite{MB0}.

\subsection{Nonresonant canonical form of $-\frac{\gamma(g)}{\beta(g)}$ by the Poincar\`e-Dulac theorem}

Hence, a sufficient condition \cite{MB0} for all the resonant terms to be absent in eq. \eqref{resft}
is that the eigenvalues of $A_0$ in nonincreasing order, $\lambda_i \geq \lambda_j$ if $i \leq j$, satisfy:
\bea \label{ein}
\lambda_i - \lambda_j \neq 2k
\eea
with $k$ a positive integer. The advantage \cite{MB0} of this sufficient condition is that it is easily verified a priori from the only knowledge of the eigenvalues of the ratio
$\frac{\gamma_0}{\beta_0}=A_0$ -- a one-loop quantity --. \par
With more effort, we can refine the sufficient condition above into a necessary and sufficient condition: If we set $A_0$ in the Jordan normal form of eq. \eqref{J},
the necessary and sufficient condition for the linear system in eq. \eqref{lin}, with $A(g)$ defined by eq. \eqref{sys}, to admit the nonresonant canonical form by a holomorphic gauge transformation:
\bea \label{eul}
A'(g)= \frac{\Lambda+N_0}{g} 
\eea
is that all of the matrix elements $(N_{2k})_{ij}$ in eq. \eqref{resft}, with $\lambda_i-\lambda_j=2k$, vanish. \par
Of course, if $\lambda_i-\lambda_j=2k$, this may only be verified by constructing iteratively (section \ref{PD}) the canonical form above. Otherwise, if $\lambda_i-\lambda_j \neq 2k$, no nonvanishing $N_k$ in eq. \eqref{resft} may occur. \par 
In both the cases above, the linear system is nonresonant, and consequently the associated operator mixing has been dubbed nonresonant in \cite{MB0}.

\subsection{Classification of operator mixing by the  Poincar\`e-Dulac theorem}

Therefore, the Poincar\`e-Dulac theorem reduces the classification \cite{MB0} of operator mixing to the four cases \cite{MB0} that we summarize below.\par

\subsubsection{(I) Nonresonant diagonalizable $\frac{\gamma_0}{\beta_0}$}

The linear system is nonresonant and $\frac{\gamma_0}{\beta_0}$ is diagonalizable.\par
For the system to be nonresonant, it is sufficient \cite{MB0} that the eigenvalues of $\frac{\gamma_0}{\beta_0}$ in nonincreasing order satisfy:
\bea \label{ein}
\lambda_i - \lambda_j \neq 2k
\eea
with  $i \leq j$ and $k$ a positive integer.\par
For $\frac{\gamma_0}{\beta_0}$ to be diagonalizable, it is sufficient that its eigenvalues are all different. \par
As we have mentioned above, for the system to be nonresonant, the necessary and sufficient condition is that in the canonical form of eq. \eqref{resft} all the resonant terms vanish.\par

\subsubsection{(II) Resonant diagonalizable $\frac{\gamma_0}{\beta_0}$}

The linear system is resonant and $\frac{\gamma_0}{\beta_0}$ is diagonalizable.\par
For the system to be resonant, a necessary condition is that, for at least two eigenvalues in nonincreasing order, it holds:
\bea \label{ein}
\lambda_i - \lambda_j = 2k
\eea
with $i < j$ and $k$ a positive integer. \par
In this case, a necessary and sufficient condition is that, correspondingly, at least one $N_{2k}$ in the canonical resonant form does not vanish.\par
The sufficient condition for $\frac{\gamma_0}{\beta_0}$ to be diagonalizable is as in the case (I).

\subsubsection{(III) Nonresonant nondiagonalizable $\frac{\gamma_0}{\beta_0}$}

The linear system is nonresonant and $\frac{\gamma_0}{\beta_0}$ is nondiagonalizable.\par
The nonresonant condition is as in the case (I). \par
The necessary condition for $\frac{\gamma_0}{\beta_0}$ to be nondiagonalizable is that at least two of its eigenvalues coincide.

\subsubsection{(IV) Resonant nondiagonalizable $\frac{\gamma_0}{\beta_0}$}

The linear system is resonant and $\frac{\gamma_0}{\beta_0}$ is nondiagonalizable.\par
The resonant condition is as in the case (II).\par
The necessary condition for $\frac{\gamma_0}{\beta_0}$ to be nondiagonalizable is as in the case (III).

\subsection{A unitarity constraint} \label{5.4}

A massless QCD-like theory is conformal \cite{ConfQCD} invariant to the leading \footnote{We assume that gauge-invariant operators are canonically normalized in such a way that the leading contribution to their $2$-point correlators starts to the order of $g^0$.}, $O(g^0)$, and next-to-leading \footnote{Implementing conformal symmetry to the order of $g^2$ requires the choice of the conformal scheme \cite{ConfQCD} that differs by a finite renormalization \cite{ConfQCD} from other perturbative schemes.}, $O(g^2)$, perturbative order, since the beta function only affects the solution of the Callan-Symanzik equation starting from the order of $g^4$.\par
We argue that, if we assume that it also is unitary in its free conformal limit at $g=0$ in the sector defined by gauge-invariant Hermitian operators, then the corresponding $\frac{\gamma_0}{\beta_0}$ should be diagonalizable. Thus, the aforementioned unitarity rules out the cases (III) and (IV). \par
The unitarity assumption above is satisfied in a massless QCD-like theory with a compact gauge group and matter satisfying the spin statistics theorem, both in Minkowskian and Euclidean space-time in the Hermitian gauge-invariant sector, as unitary gauges exist where the free limit is certainly unitary for the gluon and matter fields, and the gauge-fixing ghosts decouple in the correlators of gauge-invariant operators.\par
We demonstrate momentarily the aforementioned link between unitarity and diagonalizability of $\frac{\gamma_0}{\beta_0}$ for scalar operators in the conformal free limit. The analog argument for higher spins will appear in a forthcoming paper \cite{BB2}. \par
Firstly, if $\frac{\gamma_0}{\beta_0}$ is nondiagonalizable, to the order of $g ^2$ a logarithmic conformal field theory (logCFT) arises, which is known to be nonunitary \cite{Gur,Hog}.\par
Specifically, either in a CFT or logCFT, the $2$-point correlators of Euclidean Hermitian scalar primary conformal operators, $G_{conf}(x)$, satisfy the Callan-Symanzik equation:
\beq
\label{CSconf}
x\cdot\dfrac{\partial}{\partial x}G_{conf}(x) + \Delta G_{conf}(x) + G_{conf}(x)\Delta^T = 0
\eeq
with $\Delta$ the matrix of the conformal dimensions,
whose general solution is:
\beq
\label{G2conf}
G_{conf}(x) = \langle O(x)O(0) \rangle = e^{-\Delta \log \sqrt {x^2 \mu^2}} \mathcal{G} e^{-\Delta^T \log \sqrt {x^2 \mu^2}}
\eeq
in matrix notation, where $\mathcal{G}$ is a real symmetric matrix independent of space-time and the tensor product between repeated $O$ is understood. \par
If $\Delta$ is diagonalizable, a CFT occurs. Otherwise, if $\Delta$ is nondiagonalizable, a logCFT \cite{Gur,Hog} arises. \par
Moreover, both in a CFT and logCFT, for primary conformal operators, the operators/states correspondence holds \cite{Hog,CFT2}: 
\begin{eqnarray}
O(0)\vert 0 \rangle &=& \vert O_{in}\rangle \nonumber \\
\langle O_{out}\vert &=& \lim_{x\rightarrow \infty}\langle 0\vert e^{2\Delta\log \sqrt {x^2 \mu^2}} O(x)
\end{eqnarray}
As a consequence, the scalar product in matrix notation reads:
\begin{eqnarray}
\label{ScalarP}
\langle O_{out}\vert O_{in} \rangle &=& \lim_{x\rightarrow \infty} \langle 0 \vert e^{2\Delta\log \sqrt {x^2 \mu^2}} O(x)O(0)\vert 0 \rangle \nonumber \\
& = &\lim_{x\rightarrow \infty}  e^{2\Delta\log \sqrt {x^2 \mu^2}} e^{-\Delta \log \sqrt {x^2 \mu^2}} \mathcal{G} e^{-\Delta^T \log \sqrt {x^2 \mu^2}}  \nonumber \\
&=& \lim_{x\rightarrow \infty}  e^{\Delta\log \sqrt {x^2 \mu^2}} \mathcal{G} e^{-\Delta^T\log \sqrt {x^2 \mu^2}} 
\end{eqnarray}
In order to be well defined, the scalar product in eq. \eqref{ScalarP} must be independent of the variable $\sqrt {x^2 \mu^2}$.
Expanding the last line of eq. \eqref{ScalarP} in powers of $\log \sqrt {x^2 \mu^2}$ we get:
\begin{eqnarray}
\label{2.34}
\langle O_{out} \vert O_{in} \rangle &&= \left(I +\Delta  \mathcal{G}\log \sqrt {x^2 \mu^2} +\cdots\right)\mathcal{G}\left(I -   \mathcal{G} \Delta^T \log \sqrt {x^2 \mu^2} +\cdots\right) \nonumber \\
&&=   \mathcal{G} +\left(\Delta  \mathcal{G} -  \mathcal{G} \Delta^T \right) \log \sqrt {x^2 \mu^2} +\cdots
\end{eqnarray}
Then, the independence of the coordinates implies:
\bea \label{main}
\Delta  \mathcal{G} -  \mathcal{G} \Delta^T=0
\eea
and:
\bea
\label{2.344}
\langle O_{out} \vert O_{in} \rangle =  \mathcal{G} 
\eea
Besides, in a massless QCD-like theory, because of the existence of the perturbative expansion, it holds to the order of $g^2$:
\begin{eqnarray} \label{pert11}
\Delta(g) & = &  D \, I + g^2 \gamma_0 + \cdots \\ \nonumber
\mathcal{G}(g) & = & G_0 + g^2 G_1 + \cdots
\end{eqnarray}
in the conformal renormalization scheme \cite{ConfQCD}, with $D$ the canonical dimension of the operators $O$. Hence, expanding eq. \eqref{main} to the order of $g^2$, we obtain:
\begin{eqnarray}
\label{2.35}
\gamma_0 G_0 - G_0\gamma_0^T=0
\end{eqnarray}
Most interestingly, eq. \eqref{2.35} constrains $G_0$, which arises to the order of $g^0$, by means of $\gamma_0$, which arises to the order of $g^2$, the reason being the existence of the conformal structure to the order of $g^2$.
The consequences of eq. \eqref{2.35} follow:\par
If $\gamma_0$ is diagonalizable, by eq. \eqref{2.35} $G_0$ commutes with $\gamma_0$ in the diagonal basis and thus in any basis. Besides, $G_0$ being a real symmetric matrix, it is diagonalizable as well, and therefore $G_0$ and $\gamma_0$ are simultaneously diagonalizable.\par
This is the CFT case, where unitarity in the conformal free limit at $g=0$ requires that $G_0$ has positive eigenvalues according to eqs. \eqref{2.344} and \eqref{pert11} specialized to $g=0$.\par
If instead $\gamma_0$ is nondiagonalizable, i.e., in the logCFT case, $G_0$ has necessarily negative eigenvalues, i.e., the theory is nonunitary in its free conformal limit at $g=0$ in the gauge-invariant Hermitian sector.  \par
Indeed, if $\gamma_0$ is nondiagonalizable, eq. \eqref{main} nontrivially constrains the structure of $G_0$.
Firstly, in this case:
\beq \label{GantiDiag}
G_0 = \left(\begin{array}{ccccc}
g_1 & g_2 & g_3 & \cdots & g_n \\
g_2 & g_3 & \iddots & g_n & 0 \\
g_3 & \iddots & g_n & 0 & 0 \\
\vdots & \iddots & \vdots & \vdots & \vdots \\
g_n & 0 & \cdots & \cdots & 0
\end{array}\right)
\eeq
for some real \footnote{Not all $g_i$ may vanish, otherwise the correlator would vanish in the free conformal limit.} $g_i$ in the basis where $\gamma_0$ has the canonical Jordan form:
\bea \label{gammaJordan}
\gamma_0= \gamma_{0D} I + N_0
\eea
with $\gamma_{0D}$ the eigenvalue of the Jordan block and $N_0$ nilpotent and upper diagonal with all the nonvanishing entries equal to $1$.
Then, eq. \eqref{2.35} reads:
\be \label{N0G0}
N_{ia} G_{0aj} = G_{0ia}N^T_{aj}
\ee
with:
\beq
\label{N0}
N_{0ij} = \begin{cases}
\delta_{ij-1}  \;\;\;\;\; i = 1,\cdots,n \, ; \,\, j=2,\cdots, n   \\
 0 \;\;\;\;\;\; \;\; \;\; i = 1,\cdots,n \, ; \,\, j=1
\end{cases}
\eeq
and:
\beq
\label{N0T}
N^T_{0ij} = \begin{cases}
\delta_{i-1j} \;\;\;\;\; i = 2, \cdots,n; \,\, j = 1, \cdots, n  \\
0  \;\;\;\;\;\;\;\;\;\; i = 1 ; \,\, j = 1, \cdots, n 
\end{cases}
\eeq
Therefore, eq. \eqref{N0G0} implies:
\label{2.31d}
\bea
G_{0i+1j} &=& G_{0ij+1} \;\;\;\;\; \;\;\; \;\;\;\;  i,j = 1,\cdots,n-1 \nonumber \\
G_{0i+1n} & = & G_{0nj+1} = 0 \;\;\;\;\;    i,j = 1,\cdots,n-1 
\eea
that fixes the form of $G_0$ in eq. \eqref{GantiDiag}.\par
Moreover, by a constant gauge transformation $S$ of the form:
\bea
S= \left(\begin{array}{ccccc}
s_0 & s_1 & s_2 & \cdots & s_{n-1} \\
0 & s_0 & s_1 & \cdots & s_{n-2} \\
0 & 0 & s_0 & \cdots & s_{n-3} \\
\vdots & \vdots & \vdots & \ddots & \vdots \\
0 & 0 & 0 & \cdots & s_0
\end{array}\right)
\eea
that commutes with $N_0$,
$G_0$ transforms as \cite{BB2}:
\bea
G_0'=SG_0S^{T}
\eea
and may be set in the canonical form \cite{Hog}:
\beq \label{GantiDiag}
G'_0 = \left(\begin{array}{ccccc}
0 & 0 & 0 & \cdots & 1 \\
0 & 0 & \iddots & 1 & 0 \\
0 & \iddots & 1 & 0 & 0 \\
\vdots & \iddots & \vdots & \vdots & \vdots \\
1 & 0 & \cdots & \cdots
 & 0
\end{array}\right)
\eeq
It turns out \cite{Hog} that $G'_0$ has $[r/2]$ positive eigenvalues and $[r/2]$ negative eigenvalues, with $r$ the rank of $G_0'$. To prove the preceding statement, we observe that:
\be
G_0'^{2} = I
\ee 
Indeed, $G'_{0ij} = \delta_{i,n-j+1}$, and as a consequence:
\be
G_{0ij}'^{2} = G'_{0ia}G'_{0aj} = \delta_{i,n-a+1}\delta_{a,n-j+1}=\delta_{ij}
\ee
Hence, the eigenvalues of $G'_0$ are $\pm 1$. Moreover, the trace of $G'_0$ is either $0$ or $1$, depending on whether $r$ is even or odd respectively. \par
As a consequence, since the trace of a matrix is the sum of its eigenvalues, if $r$ is even, $G'_0$ has $r/2$ positive eigenvalues and $r/2$ negative eigenvalues, otherwise, if $r$ is odd, $G'_0$ has $r/2 + 1$ positive eigenvalues and $r/2$ negative eigenvalues.\par
By summarizing, the key point of the argument above is that the nondiagonalizability of $\gamma_0$ and the existence of the conformal structure to the order of $g^2$
determine the structure of $G_0$ that controls the scalar product in the free conformal limit, in such a way that the free conformal limit is nonunitary if $\gamma_0$ is nondiagonalizable. \par
Finally, we may extend the perturbative argument above -- about the existence of the scalar product to the order of $g^2$ -- to all orders of perturbation theory,
by considering a massless QCD-like theory at its conformal Wilson-Fisher fixed point $g_*$, with $\beta(g_*,\epsilon)=-g_* \epsilon+ \beta(g)=0$, introduced in \cite{Braun1,Braun2} to perform higher-loop computations in dimensional regularization -- in $d=4-2 \epsilon$ dimensions -- of the anomalous-dimension matrices in massless QCD. \par
Indeed, the anomalous-dimension matrix $\gamma(g_*)$ at the fixed point has the same coefficients \cite{Braun1,Braun2} -- as a series in $g_*$ -- as the anomalous-dimension matrix $\gamma(g)$ -- as a series in $g$ -- and specifically the same $\gamma_0$. Moreover, since the theory is conformal to all perturbative orders at the fixed point, the associated scalar product exists to all orders in $g_*$. \par
Either way, the perturbative conformal symmetry to the order of $g^2$ or the conformal symmetry to all orders at the aforementioned Wilson-Fisher fixed point, and the lowest-order unitarity, rule out the cases (III) and (IV) of operator mixing in the gauge-invariant Hermitian sector of a massless QCD-like theory. The statement above does not necessarily apply to operators outside the gauge-invariant sector, whose correlators may be affected by the mixing with the ghost sector, which need not to be unitary.

\subsection{A condensed proof of the Poincar\`e-Dulac theorem} \label{PD}

We provide a condensed proof of (the linear version of) the Poincar\`e-Dulac theorem following \cite{PD1}.\par

\emph{Poincar\`e-Dulac theorem}:\par

The most general linear system with a Fuchsian singularity at $g=0$, 
where the meromorphic connection $A(g)$ admits the (formal) expansion:
\bea \label{sys2}
A(g)= \frac{1}{g} \left(A_0 + \sum^{\infty}_ {n=1} A_{n} g^{n} \right)
\eea
may be set, by a (formal) holomorphic invertible gauge transformation, in the Poincar\`e-Dulac-Levelt normal form \footnote{In the present paper, we refer to it as the resonant canonical form.}:
\bea \label{canres1}
A'(g)= \frac{1}{g} \left(\Lambda+N_0 + \sum_ {k=1} N_{k} g^{k} \right)
\eea
where $\Lambda+N_0$ is the Jordan normal form of $A_0$, its eigenvalues $\text{diag}(\lambda_1, \lambda_2 \cdots )=\Lambda $ are in nonincreasing order
$\lambda_1 \geq \lambda_2 \geq \cdots$, $N_0$ is nilpotent and upper triangular,
and the nilpotent upper triangular matrices $N_k$ satisfy:
\bea \label{gr}
 g^{\Lambda} N_k g^{-\Lambda} = g^k N_k
 \eea
for $k=1,2,\cdots$, i.e., the only possibly nonvanishing matrix elements, $(N_k)_{ij}$, of the $N_k$ are associated to the resonant eigenvalues:
\bea
\lambda_i-\lambda_j=k
\eea
with $i < j$ and $k$ a positive integer.\par
Incidentally, also $g^{\Lambda} N_0 g^{-\Lambda} = N_0$, since $[N_0,\Lambda]=0$ by the Jordan normal form of $A_0$. \par
Of course, if either the eigenvalues are nonresonant or the resonant matrix coefficients $N_k$ -- associated to the resonant eigenvalues -- vanish, the linear system collapses \cite{PD1} into the Euler form \footnote{In the present paper, we refer to it as the nonresonant canonical form.}: 
\bea \label{canres2}
A'(g)= \frac{1}{g} \left(\Lambda+N_0 \right)
\eea
We only report the key aspects of the proof, leaving more details to \cite{PD1}. \par
\emph{Proof}:\par
The demonstration proceeds by induction on $k=1,2, \cdots$ by proving that, once $A_0$ and the first $k-1$ matrix coefficients, $A_1, \cdots, A_{k-1}$, have been set in the Poincar\`e-Dulac-Levelt normal form above, a holomorphic gauge transformation exists that leaves them invariant and also puts the $k$-th coefficient, $A_{k}$, in normal form. \par
The step $0$ of the induction consists just in putting $A_0$ in Jordan normal form -- with eigenvalues in nonincreasing order and $N_0$ upper triangular, as in the statement of the theorem -- by a global (i.e., constant) gauge transformation. \par
At the $k$-th step, we choose the holomorphic gauge transformation of the form: 
\bea \label{gt00}
S_k(g)=1+ g^k H_k
\eea
with $H_k$ a matrix to be found momentarily. Its inverse is:
\bea
S^{-1}_k(g)= (1+ g^k H_k)^{-1} = 1- g^k H_k + \cdots
\eea
where the dots represent terms of order higher than $g^{k}$.
The gauge action of $S_k(g)$ on the connection $A(g)$ furnishes:
\bea \label{ind}
A'(g) &= & k g^{k-1} H_k ( 1+ g^k H_k)^{-1}+  (1+ g^k H_k) A(g)( 1 + g^k H_k)^{-1} \nonumber \\
&= & k g^{k-1} H_k ( 1 + g^k H_k)^{-1} +  (1+ g^k H_k)  \frac{1}{g} \left(A_0 + \sum^{\infty}_ {n=1} A_{n} g^{n} \right)  ( 1 + g^k H_k)^{-1} \nonumber \\
&= & k g^{k-1} H_k ( 1- \cdots) +  (1+ g^k H_k)  \frac{1}{g} \left(A_0 + \sum^{\infty}_ {n=1} A_{n} g^{n} \right)  ( 1- g^k H_k+\cdots) \nonumber \\
&= & k g^{k-1} H_k  +    \frac{1}{g} \left(A_0 + \sum^k_ {n=1} A_{n} g^{n} \right) + g^{k-1} (H_kA_0-A_0H_k) + \cdots\nonumber \\
&= &  g^{k-1} (k H_k + H_k A_0 - A_0 H_k)  + A_{k-1}(g) + g^{k-1} A_k+ \cdots\nonumber \\
\eea
where we have skipped in the dots all the terms that contribute to an order higher than $g^{k-1}$, and we have put:
\bea
A_{k-1}(g) =  \frac{1}{g} \left(A_0 + \sum^{k-1}_ {n=1} A_{n} g^{n} \right)
\eea
that is the part of $A(g)$ that is not affected by the gauge transformation $S_k(g)$, and thus it verifies the hypotheses of the induction.\par
Therefore, by eq. \eqref{ind} the $k$-th matrix coefficient, $A_k$, may be eliminated from the expansion of $A'(g)$ to the order of $g^{k-1}$ provided that an $H_k$ exists such that:
\bea \label{Hkeq}
A_k+(k H_k + H_k A_0 - A_0 H_k)= A_k+ (k-ad_{A_0}) H_k=0
\eea
with $ad_{A_0}Y=[A_0,Y]$.
If the inverse of $ad_{A_0}-k$ exists, the unique solution for $H_k$ is:
\bea
H_k=(ad_{A_0}-k)^{-1} A_k
\eea
Therefore, the only matrix coefficients that may not be removed from the expansion of $A'(g)$ at the $k$-th step of the induction belong to the subspace where $ad_{A_0}-k$ is not invertible.\par
Hence, we should demonstrate that, for $k$ positive, the elements $Y_k$ of the aforementioned subspace satisfy the condition in eq. \eqref{gr} for $N_k$, according to the statement of the theorem.\par
To understand what is going on, it is convenient to suppose initially \cite{MB0} that $N_0=0$, i.e., that $A_0$ is diagonalizable. \par
In this case, $ad_{A_0}-k=ad_{\Lambda}-k$, as a linear operator that acts on matrices, is diagonal with eigenvalues $\lambda_{i}-\lambda_{j}-k$ and the matrices $E_{ij}$, whose only nonvanishing entries are $(E_{ij})_{ij}$, as eigenvectors. Moreover, $ad_{\Lambda}-k$ is invertible if and only if its kernel only contains the zero matrix. \par
The eigenvectors $E_{ij}$, normalized in such a way that $(E_{ij})_{ij}=1$, form an orthonormal basis for the matrices: $\langle E_{ij}|E_{i'j'} \rangle= \delta_{ii'} \delta_{jj'}$, with $\langle A  |B\rangle=\Tr(\bar{A} B)$ and $\bar{A}$ the adjoint of the matrix $A$.  \par
Thus, $E_{ij}$ belongs to the kernel of $ad_{\Lambda}-k$ if and only if $\lambda_{i} - \lambda_{j}-k=0$ and $i<j$, as $k$ is a positive integer. \par
As a consequence, the $E_{ij}$ in the kernel satisfy eq. \eqref{gr}, according to the statement of the theorem:
 \bea
 g^{\Lambda} E_{ij} g^{-\Lambda}= g^{\lambda_{i}-\lambda_{j}} E_{ij}= g^k E_{ij}
 \eea
Now we suppose that $N_0$ does not vanish, i.e., $A_0$ is nondiagonalizable. \par
Hence, $A_0$ admits a canonical Jordan form as in the statement of the theorem. \par
The key point is that now $ad_{A_0}-k$, as a linear operator that acts on matrices, is lower triangular for the following ordering of the matrix basis.  \par
We may choose an increasing sequence, $\text{diag} (q_1,q_2, \cdots)=Q$, of rationally independent weights, $q_i$, \cite{PD1} in such a way that the corresponding weight for $E_{ij}$ is $q_j-q_i$ -- computed via $g^{-Q} E_{ij} g^{Q} = g^{q_j-q_i} E_{ij}$ --. \par
Thus, we may order our basis in such a way that the sequence of basis vectors $E_l$ with $l=1,2, \cdots$ coincides with the following sequence of the $E_{ij}$ ordered with nondecreasing weights: The $E_{ij}$ for $i \neq j$ with strictly increasing weights, and the $E_{ii}$ -- that have weight $0$ -- with $i$ increasing.\par
The action of $ad_{\Lambda}$ on the basis leaves the weights of the $E_{ij}$ with $i \neq j$ invariant, and sends to zero the $E_{ii}$, in such a way that the action of $ad_{\Lambda}$ is diagonal on the entire basis. \par
Instead, the action of $ad_{N_0}$ on the entire basis produces a linear combination of terms with strictly increased weights, since $N_0$ is upper triangular and, therefore, it is the sum of terms with positive weights, and for each of these terms the commutator with any $E_{ij}$ strictly increases the weights.\par 
Moreover:
\bea
(ad_{\Lambda+N_0}-k) E_l =  E_h  \langle E_h |(ad_{\Lambda+N_0}-k) E_l \rangle=   E_h (ad_{\Lambda+N_0}-k)_{hl} 
\eea
where the sum on the index $h$ is understood.
Hence, with this ordering of the basis, the matrix:
\bea 
(ad_{\Lambda+N_0}-k)_{hl}=\langle E_h |(ad_{\Lambda+N_0}-k)E_l \rangle
\eea
is lower triangular \cite{PD1} and its eigenvalues coincide with the eigenvalues of $ad_{\Lambda}-k$.\par
Now $ad_{\Lambda+N_0}-k$ is not invertible if and only if at least one of its eigenvalues vanishes. But its eigenvalues coincide with the eigenvalues of
$ad_{\Lambda}-k$. Therefore, $ad_{\Lambda+N_0}-k$ is invertible on the orthogonal complement of the kernel of $ad_{\Lambda}-k$, as it is for $ad_{\Lambda}-k$.\par
Hence, every matrix coefficient $A_k$ orthogonal to the kernel of $ad_{\Lambda}-k$ may be removed from $A'(g)$, as in the diagonalizable case with $N_0=0$ above.\par
Obviously, the resonant matrix coefficients $N_k$ are finite in number, because there are only a finite number of differences of the eigenvalues.\par
As consequence, from a certain point on, all the remaining terms in the expansion of $A'(g)$ may be removed, because they belong to the orthogonal complement of the kernel of $ad_{A_0}-k$, and the proof is complete.\par

\subsection{Fundamental solution of the linear system} \label{F}

A fundamental solution \cite{PD1} of the linear system in eq. \eqref{lin} in the canonical resonant form of eq. \eqref{canres1} is:
\bea \label{fsol}
X(g)= g^{\Lambda} g^{N}
\eea
with $N=N_0+\sum_{k=1} N_{k}$, as we verify by direct computation \cite{PD1}:
\bea
\frac{\partial{X(g)}}{{\partial g}} X^{-1}(g)&=&  g^{\Lambda}    \frac{\Lambda+N}{g} g^{N} g^{-N} g^{-\Lambda} \nonumber \\
&=&  g^{\Lambda}    \frac{\Lambda+N}{g} g^{-\Lambda} \nonumber \\
&=&  \frac{\Lambda+g^{\Lambda} N g^{-\Lambda}}{g}  \nonumber \\
&=&  \frac{\Lambda+N_0 + \sum_ {k=1} N_{k} g^{k}}{g}  \nonumber \\
&=&A'(g) \nonumber \\
\eea
Moreover, $X(g)$ may be computed in a closed form, since the expansion of $g^{N}$ in powers of $\log g$ terminates because $N$ is nilpotent.\par
Correspondingly, the solution $X(g)X^{-1}(g_0)$ of eq. \eqref{lin} in the canonical form of eq. \eqref{canres1} that reduces to the identity at $g=g_0$ may be computed in a closed form as well:
\bea
X(g)X^{-1}(g_0)&=&g^{\Lambda} g^{N} g^{-N}_0 g^{-\Lambda}_0 \nonumber \\
&=&g^{\Lambda} \left(\frac{g}{g_0}\right)^N g_0^{-\Lambda} \nonumber \\
&=&\left(\frac{g}{g_0}\right)^{\Lambda}    g^{\Lambda}_0  \left(\frac{g}{g_0}\right)^N g^{-\Lambda}_0 \nonumber \\
&=&\left(\frac{g}{g_0}\right)^{\Lambda}    g^{\Lambda}_0 e^{N \log \frac{g}{g_0} } g^{-\Lambda}_0 \nonumber \\
&=&\left(\frac{g}{g_0}\right)^{\Lambda}     e^{g^{\Lambda}_0 N g^{-\Lambda}_0 \log \frac{g}{g_0} }  \nonumber \\
&=&\left(\frac{g}{g_0}\right)^{\Lambda}     e^{\sum_{k=0} g^{k}_0 N_{k} \log \frac{g}{g_0} }  \nonumber \\
\eea

\subsection{Solution for $Z(x,\mu)$} \label{ZZ}

Therefore, the solution of eq. \eqref{lin} in the canonical resonant form of eq. \eqref{resft} that reduces to the identity for $g(x)=g(\mu)$ in a massless QCD-like theory is:
\bea \label{zz}
Z(x,\mu)&=&g^{\Lambda}(\mu) g^{N}(\mu) g^{-N}(x) g^{-\Lambda}(x) \nonumber \\
&=&g^{\Lambda}(\mu) \left(\frac{g(\mu)}{g(x)}\right)^N g^{-\Lambda}(x) \nonumber \\
&=&\left(\frac{g(\mu)}{g(x)}\right)^{\Lambda}    g^{\Lambda}(x)  \left(\frac{g(\mu)}{g(x)}\right)^N g^{-\Lambda}(x) \nonumber \\
&=&\left(\frac{g(\mu)}{g(x)}\right)^{\Lambda}    g^{\Lambda}(x) e^{N \log \frac{g(\mu)}{g(x)} } g^{-\Lambda}(x) \nonumber \\
&=&\left(\frac{g(\mu)}{g(x)}\right)^{\Lambda}     e^{g^{\Lambda}(x)Ng^{-\Lambda}(x) \log \frac{g(\mu)}{g(x)} }  \nonumber \\
&=&\left(\frac{g(\mu)}{g(x)}\right)^{\Lambda}     e^{\sum_{k=0} g^{2k}(x) N_{2k} \log \frac{g(\mu)}{g(x)} }  \nonumber \\
\eea
where \cite{MB0} $g(\mu)$ and $g(x)$ are short notations for the running coupling at the corresponding scales, $g(\frac{\mu}{\Lambda_{RGI}})$ and $g(x \Lambda_{RGI})$, and:
\begin{equation}
\label{1.12}
g^2(x \Lambda_{RGI})  \sim \dfrac{1}{\beta_0\log(\frac{1}{x^2 \Lambda_{RGI}^2})} \left(1-\dfrac{\beta_1}{\beta_0^2} \dfrac{\log\log(\frac{1}{x^2 \Lambda_{RGI}^2})}{\log(\frac{1}{x^2 \Lambda_{RGI}^2})}\right)
\end{equation}

\subsection{UV asymptotics of $Z(x,\mu)$}

In the cases (II),(III) and (IV), as the canonical form of $Z(x,\mu)$ is nondiagonal, its UV asymptotics is intrinsically different from the diagonal case (I). \par
Indeed, by eq. \eqref{zz} $Z(x,\mu)$ may be factorized into the exponential of a linear combination of upper triangular nilpotent matrices with coefficients that asymptotically in the UV are powers of logs of the running coupling, i.e., powers of loglogs of the coordinates, and a diagonal matrix as in the nonresonant diagonal case (I): 
\bea \label{LL00}
Z(x,\mu)&=&\left(\frac{g(\mu)}{g(x)}\right)^{\Lambda}     e^{\sum_{k=0} g^{2k}(x) N_{2k} \log \frac{g(\mu)}{g(x)} }  \nonumber \\
&=&\left(\frac{g(\mu)}{g(x)}\right)^{\Lambda}     e^{\sum_{k=0} g^{2k}(x) N_{2k} \log \frac{g(\mu)}{g(x)} } \left(\frac{g(\mu)}{g(x)}\right)^{-\Lambda} \left(\frac{g(\mu)}{g(x)}\right)^{\Lambda} \nonumber \\
&=&   e^{\sum_{k=0} g^{2k}(\mu) N_{2k} \log \frac{g(\mu)}{g(x)} }  \left(\frac{g(\mu)}{g(x)}\right)^{\Lambda} 
\eea
This is the closest analog of logCFTs that may occur in asymptotically free theories.\par
Moreover, some subtleties arise in computing the UV asymptotics of $Z(x,\mu)$, since it follows from eq. \eqref{LL00} that the factorization of $Z(x,\mu)$ actually depends on the order of the factors, in such a way that:
\bea \label{LL}
\left(\frac{g(\mu)}{g(x)}\right)^{-\Lambda}  Z(x,\mu) &=&   e^{\sum_{k=0} g^{2k}(x) N_{2k} \log \frac{g(\mu)}{g(x)} }  
\eea
but:
\bea \label{RR}
Z(x,\mu)  \left(\frac{g(\mu)}{g(x)}\right)^{-\Lambda}&=&   e^{\sum_{k=0} g^{2k}(\mu) N_{2k} \log \frac{g(\mu)}{g(x)} } 
\eea
Therefore, the limit for $x \rightarrow 0$ of eqs. \eqref{LL} and \eqref{RR} in general does not coincide. \par
Specifically, in the case (II), i.e., for $N_0=0$, the limit is the identity $I$ for eq. \eqref{LL}, but it is not finite for eq. \eqref{RR}.

\section{Three examples for the mixing of two operators revisited by the Poincar\`e-Dulac theorem} \label{6}

For completeness, we verify that the elementary computation in section \ref{5} coincides with the solution of the linear system in canonical form according to the Poincar\`e-Dulac theorem.

\subsection{Nonresonant diagonalizable $\frac{\gamma_0}{\beta_0}$}

$N_0=0$, because $\frac{\gamma_0}{\beta_0}$ is diagonal.
Moreover:
\be
\Lambda=
\left(\begin{array}{cc}
\lambda_1 &  0  \\
0 &  \lambda_2
\end{array}\right)
\ee
and:
\bea
N_{2k}=0
\eea
for $k=1,2, \cdots$, because the system is nonresonant. Therefore, by eq. \eqref{zz} we obtain:
\bea
Z(x,\mu)&=&\left(\frac{g(\mu)}{g(x)}\right)^{\Lambda} \nonumber \\
              &=&  \begin{scriptsize}\left(
\begin{array}{cc}
 \left(\dfrac{g(\mu)}{g(x)}\right)^{\lambda_1} & 0 \\
 0 &  \left(\dfrac{g(\mu)}{g(x)}\right)^{\lambda_2} \\
\end{array}
\right)
\end{scriptsize}
\eea
that matches eq. \eqref{ab} up to a holomorphic gauge transformation.

\subsection{Resonant diagonalizable $\frac{\gamma_0}{\beta_0}$}

$N_0=0$, because $\frac{\gamma_0}{\beta_0}$ is diagonal. Moreover:
\bea
\Lambda=
\left(\begin{array}{cc}
\lambda_1 &  0  \\
0 &  \lambda_2
\end{array}\right)
\eea
and:
\be
N=N_{2k}=
\left(\begin{array}{cc}
0 &  \nu_{12}  \\
0 & 0
\end{array}\right)
\ee
with $\lambda_1-\lambda_2=2k$, since the system is resonant.
Therefore, by eq. \eqref{zz} we obtain:
\bea
Z(x,\mu) 
&& = \begin{scriptsize}
\left(	
\begin{array}{cc}
 \left(\dfrac{g(\mu)}{g(x)}\right)^{\lambda_1} & 0 \\
 0 & \left(\dfrac{g(\mu)}{g(x)}\right)^{\lambda_2}
\end{array}
\right) \end{scriptsize}\left(I + \left(\begin{array}{cc}
0 & \nu_{12}g^{2k}(x) \log \dfrac{g(\mu)}{g(x)} \\
0 & 0
\end{array}\right)\right) \nonumber \\
&& = \begin{scriptsize}\left(
\begin{array}{cc}
 \left(\dfrac{g(\mu)}{g(x)}\right)^{\lambda_1} & \nu_{12}g^{2k}(x)\left(\dfrac{g(\mu)}{g(x)}\right)^{\lambda_1}\log \dfrac{g(\mu)}{g(x)} \\
 0 & \left(\dfrac{g(\mu)}{g(x)}\right)^{\lambda_2}
\end{array}
\right)
\end{scriptsize} \nn \\
\eea 
that matches eq. \eqref{zsol2}.

\subsection{Nonresonant nondiagonalizable $\frac{\gamma_0}{\beta_0}$}

$\frac{\gamma_0}{\beta_0}$ is not diagonalizable. Hence:
\bea
\Lambda=
\left(\begin{array}{cc}
\lambda &  0  \\
0 &  \lambda
\end{array}\right)
\eea
and:
\be
N=N_{0}=
\left(\begin{array}{cc}
0 &  \nu_{12}  \\
0 & 0
\end{array}\right)
\ee
Therefore, by eq. \eqref{zz} we obtain:
\bea
Z(x,\mu) 
&& = \begin{scriptsize}
\left(	
\begin{array}{cc}
 \left(\dfrac{g(\mu)}{g(x)}\right)^{\lambda} & 0 \\
 0 & \left(\dfrac{g(\mu)}{g(x)}\right)^{\lambda}
\end{array}
\right) \end{scriptsize}\left(I + \left(\begin{array}{cc}
0 & \nu_{12} \log \dfrac{g(\mu)}{g(x)} \\
0 & 0
\end{array}\right)\right)
\nonumber \\
&& = \begin{scriptsize}\left(
\begin{array}{cc}
 \left(\dfrac{g(\mu)}{g(x)}\right)^{\lambda} & \nu_{12}\left(\dfrac{g(\mu)}{g(x)}\right)^{\lambda}\log \dfrac{g(\mu)}{g(x)} \\
 0 & \left(\dfrac{g(\mu)}{g(x)}\right)^{\lambda}
\end{array}
\right)
\end{scriptsize} \nn \\
\eea 
that matches eq. \eqref{Zsol}.

\section{A physical realization of the cases (II) and (I)} \label{7}

\subsection{Flavor-changing four-quarks operators in SU($N$) QCD with $N_f=N$ flavors of quarks}

We work out a physical realization of the case (II): The operator mixing of flavor-changing four-quarks operators computed in \cite{Buras3}. Specifically, we consider the two sets of operators \cite{Buras3}:
\bea \label{VLR}
Q_1^{VLR} &  = & \left(\bar{s}^{\alpha} \gamma_{\mu} P_L d^{\beta}\right) \left(\bar{u}^{\beta} \gamma^{\mu} P_R c^{\alpha}\right) \nn \\
Q_2^{VLR} &  = & \left(\bar{s}^{\alpha} \gamma_{\mu} P_L d^{\alpha}\right) \left(\bar{u}^{\beta} \gamma^{\mu} P_R c^{\beta}\right)
\eea
and:
\bea \label{SLR}
Q_1^{SLR} &  = & \left(\bar{s}^{\alpha} P_L d^{\beta}\right) \left(\bar{u}^{\beta}  P_R c^{\alpha}\right) \nn \\
Q_2^{SLR} &  = & \left(\bar{s}^{\alpha}  P_L d^{\alpha}\right) \left(\bar{u}^{\beta}  P_R c^{\beta}\right)
\eea
where $\bar{s}$, $d$, $\bar{u}$ and $c$ are quarks operators, and $P_{L,R} = \frac{1}{2}\left(1 \mp \gamma_5\right)$. 
In \cite{Buras3} the anomalous-dimension matrices have been computed to the order of $g^4$:
\be
\gamma(g) = g^2 \gamma_0 + g^4 \gamma_1 + \cdots
\ee
where $\gamma_0$ and $\gamma_1$ are:
\bea
\gamma_0^{VLR}  & = & \frac{1}{(4\pi)^2}\left( \begin{array}{cc} -6  + \frac{6}{N^2} & 0 \\ -\frac{6}{N} & \frac{6}{N^2} \end{array}\right) \nn \\
\gamma_1^{VLR} &=&  \frac{1}{(4\pi)^4}\left( \begin{array}{ccc}
-\f{203}{6} N +\f{479}{6N} + \f{15}{2N^3} + \f{10}{3} N_f -\f{22}{3N^2} N_f &~~~&
-\f{71}{2} - \f{18}{N^2} +4\frac{N_f}{N} \\[1mm] 
- \f{100}{3} + \f{3}{N^2} + \f{22}{3} \frac{N_f}{N} &&
\f{137}{6N} + \f{15}{2N^3} - \f{22}{3N^2} N_f
\end{array} \right) \nn \\
\eea
for the VLR operators, and:
\bea
\gamma_0^{SLR} & = &  \frac{1}{(4\pi)^2}\left( \begin{array}{cc} \frac{6}{N^2} & -\frac{6}{N} \\ 0 & -6 + \frac{6}{N^2} \end{array}\right)\nn  \\
\gamma_1^{SLR} &=&  \frac{1}{(4\pi)^4}\left( \begin{array}{ccc}
\f{137}{6N} + \f{15}{2N^3} - \f{22}{3N^2} N_f &~~~&
-\f{100}{3} + \f{3}{N^2} + \f{22}{3} \frac{N_f}{N} \\[1mm] 
- \f{71}{2} - \f{18}{N^2} + 4\frac{N_f}{N} &&
-\f{203}{6} N +\f{479}{6N} + \f{15}{2N^3} + \f{10}{3} N_f -\f{22}{3N^2} N_f
\end{array} \right) \nn \\
\eea
for the SLR operators, with $N$ and $N_f$ the number of colors and flavors respectively.\par
The eigenvalues of $\gamma_0^{VLR}$ in nonincreasing order are:
\bea
\lambda_1^{VLR} & = & \frac{1}{(4\pi)^2} \frac{6}{N^2} \nn \\
\lambda_2^{VLR} & = & \frac{1}{(4\pi)^2} 6\left(-1 +\frac{1}{N^2}\right)
\eea
that coincide with the eigenvalues of $\gamma_0^{SLR}$:
\bea
\lambda_1^{SLR} & = &  \frac{1}{(4\pi)^2} \frac{6}{N^2} \nn \\
\lambda_2^{SLR} & = & \frac{1}{(4\pi)^2} 6\left(-1 +\frac{1}{N^2}\right)
\eea
We set $N_f = N$, in such a way that $\beta_0$ and $\beta_1$ read respectively:
\bea
\beta_0 &=& \frac{1}{(4\pi)^2}\left(\frac{11}{3} - \frac{2}{3}\frac{N_f}{N}\right) = \frac{3}{(4\pi)^2} \nonumber \\
\beta_1 &=& \frac{1}{(4\pi)^4}\left(\frac{34}{3} - \frac{13}{3}\frac{N_f}{N} + \frac{N_f}{N^3}\right) = \frac{1}{(4\pi)^4}\left(7 + \frac{1}{N^2}\right)
\eea
Therefore, the differences of the eigenvalues satisfy the resonant condition in eq. \eqref{resex2} with $k=1$:
\be
\frac{\lambda_1^{VLR}}{\beta_0}-\frac{\lambda_2^{VLR}}{\beta_0} = 2
\ee
\be
\frac{\lambda_1^{SLR}}{\beta_0}-\frac{\lambda_2^{SLR}}{\beta_0} = 2
\ee
thus realizing in a physical theory the case (II). \par
We construct the holomorphic gauge transformations that bring the corresponding connections:
\be \label{gVLRconn}
-\frac{\gamma^{VLR}(g)}{\beta(g)}  =  \frac{1}{g}\left(\frac{\gamma_{0D}^{VLR}}{\beta_0} + g^2 \frac{\beta_0 \gamma_{1D}^{VLR} - \beta_1 \gamma_{0D}^{VLR}}{\beta_0^2}\right) + \cdots
\ee
and:
\be  \label{gSLRconn}
-\frac{\gamma^{SLR}(g)}{\beta(g)}  =  \frac{1}{g}\left(\frac{\gamma_{0D}^{SLR}}{\beta_0} + g^2 \frac{\beta_0 \gamma_{1D}^{SLR} - \beta_1 \gamma_{0D}^{SLR}}{\beta_0^2}\right) + \cdots
\ee
in the Poincar\`e-Dulac-Levelt normal form of eq. \eqref{canres1}. \par
Firstly, we get the bases of operators where $\gamma_0^{VLR}$ and $\gamma_0^{SLR}$ are diagonal, by means of the global gauge transformations:
\be
S^{VLR}_0  = \left(
\begin{array}{cc}
 -\frac{1}{N} & 1 \\
 \frac{1}{N} & 0 \\
\end{array}
\right)
\ee
and:
\be
S^{SLR}_0  =  \left(
\begin{array}{cc}
 1 & -\frac{1}{N} \\
 0 & 1 \\
\end{array}
\right)
\ee
respectively. Correspondingly:
\bea
\gamma_{0D}^{VLR} & = & \frac{1}{(4\pi)^2}\left(
\begin{array}{cc}
\frac{6}{N^2}   & 0 \\
 0 & \frac{6}{N^2}-6\\
\end{array}
\right) \nn \\
\gamma_{1D}^{VLR} & = &  \frac{1}{(4\pi)^4}\left(
\begin{array}{cc}
 \frac{51}{2 N^3}+\frac{47}{N} & \frac{18}{N^3}+\frac{9 N}{2}-\frac{45}{2 N} \\
-\frac{18}{N^3}-\frac{63}{2 N} & \frac{41}{N}-\frac{21}{2 N^3}-\frac{61 N}{2} \\
\end{array}
\right)
\eea
and:
\bea
\gamma_{0D}^{SLR} & = & \frac{1}{(4\pi)^2}\left(
\begin{array}{cc}
 \frac{6}{N^2} & 0 \\
 0 & \frac{6}{N^2}-6 \\
\end{array}
\right) \nn \\
\gamma_{1D}^{SLR} & = &  \frac{1}{(4\pi)^4}\left(
\begin{array}{cc}
 \frac{51}{2 N^3}+\frac{47}{N} & \frac{18}{N^4}-\frac{45}{2 N^2}+\frac{9}{2} \\
 -\frac{18}{N^2}-\frac{63}{2} & \frac{41}{N}-\frac{21}{2 N^3}-\frac{61 N}{2} \\
\end{array}
\right)
\eea
Secondly, we choose the gauge transformations in eq. \eqref{gt00} for $k=2$:
\be
S_2^{VLR}(g)  =  I + g^2 H^{VLR}_2 = \left(\begin{array}{cc} 1 & 0 \\ 0 & 1 \end{array}\right) + g^2 \left(\begin{array}{cc} a^{VLR}_{11} & 0 \\ a^{VLR}_{21} & a^{VLR}_{22} \end{array}\right)
\ee
and:
\be
S_2^{SLR}(g)  =  I + g^2 H^{SLR}_2 = \left(\begin{array}{cc} 1 & 0 \\ 0 & 1 \end{array}\right) + g^2 \left(\begin{array}{cc} a^{SLR}_{11} & 0\\ a^{SLR}_{21} & a^{SLR}_{22} \end{array}\right)
\ee
respectively, by requiring that the only terms that do not vanish in the gauge-transformed eqs. \eqref{gVLRconn} and \eqref{gSLRconn} are the resonant ones:
\be
A^{VLR}_2 + (2\, I - \operatorname{ad}_{A^{VLR}_0})H^{VLR}_2 = \frac{1}{(4\pi)^2}\left(
\begin{array}{cc}
0  & \frac{6}{N^3}+\frac{3 N}{2}-\frac{15}{2 N} \\
 0 & 0 \\
\end{array}
\right)
\ee
and:
\be
A^{SLR}_2 + (2\, I - \operatorname{ad}_{A^{SLR}_0})H^{SLR}_2 = \frac{1}{(4\pi)^2}\left(
\begin{array}{cc}
0  &  \frac{6}{N^4}-\frac{15}{2 N^2}+\frac{3}{2} \\
 0 & 0 \\
\end{array}
\right)
\ee
respectively, where:
\bea
A^{VLR}_0 & = & \frac{\gamma_{0D}^{VLR}}{\beta_0} = \left(
\begin{array}{cc}
\frac{2}{N^2}  & 0 \\
 0 & \frac{2}{N^2}-2 \\
\end{array}
\right) \nn \\
A^{VLR}_2 & = &  \frac{\beta_0 \gamma_{1D}^{VLR} - \beta_1 \gamma_{0D}^{VLR}}{\beta_0^2} \nn \\ 
& = &\frac{1}{(4\pi)^2}\left(
\begin{array}{cc}
 -\frac{2}{3 N^4}+\frac{17}{2 N^3}-\frac{14}{3 N^2}+\frac{47}{3 N} & \frac{6}{N^3}+\frac{3 N}{2}-\frac{15}{2 N} \\
 -\frac{6}{N^3}-\frac{21}{2 N} & -\frac{2}{3 N^4}-\frac{7}{2 N^3}-\frac{4}{N^2}-\frac{61 N}{6}+\frac{41}{3 N}+\frac{14}{3} \\
\end{array}
\right) \nn \\
\eea
and:
\bea
A^{SLR}_0 &  = &  \frac{\gamma_{0D}^{SLR}}{\beta_0} = \left(
\begin{array}{cc}
 \frac{2}{N^2} & 0 \\
 0 & \frac{2}{N^2}-2 \\
\end{array}
\right) \nn \\
A^{SLR}_2 & = &  \frac{\beta_0 \gamma_{1D}^{SLR} - \beta_1 \gamma_{0D}^{SLR}}{\beta_0^2} \nn \\ 
& = & \frac{1}{(4\pi)^2}\left(
\begin{array}{cc}
 -\frac{2}{3 N^4}+\frac{17}{2 N^3}-\frac{14}{3 N^2}+\frac{47}{3 N} & \frac{6}{N^4}-\frac{15}{2 N^2}+\frac{3}{2} \\
 -\frac{6}{N^2}-\frac{21}{2} & -\frac{2}{3 N^4}-\frac{7}{2 N^3}-\frac{4}{N^2}-\frac{61 N}{6}+\frac{41}{3 N}+\frac{14}{3} \\
\end{array}
\right) \nn \\
\eea
Therefore:
\bea \label{H2eq}
A^{VLR}_2 + (2\, I - \operatorname{ad}_{A^{VLR}_0})H^{VLR}_2 
& = & \left(
\begin{array}{cc}
 2 a_{11}^{VLR} + c_1^{VLR} & \frac{1}{(4\pi)^2}\left(\frac{6}{N^3}+\frac{3 N}{2}-\frac{15}{2 N}\right) \\
 4 a_{21}^{VLR}+c_2^{VLR} & 2 a_{22}^{VLR}+c_3^{VLR} \\
\end{array}
\right) \nn \\
 &=&  \frac{1}{(4\pi)^2}\left(
\begin{array}{cc}
0  & \frac{6}{N^3}+\frac{3 N}{2}-\frac{15}{2 N} \\
 0 & 0 \\
\end{array}
\right) 
\eea
with:
\bea
c_1^{VLR} & = & \frac{1}{(4\pi)^2}\left(-\frac{2}{3 N^4}+\frac{17}{2 N^3}-\frac{14}{3 N^2}+\frac{47}{3 N}\right)\nn \\
c_2^{VLR} & = & \frac{1}{(4\pi)^2}\left(-\frac{6}{N^3}-\frac{21}{2 N}\right) \nn \\
c_3^{VLR} & = & \frac{1}{(4\pi)^2}\left(\frac{14}{3}-\frac{2}{3 N^4}-\frac{7}{2 N^3}-\frac{4}{N^2}-\frac{61 N}{6}+\frac{41}{3 N}\right)
\eea
and:
\bea
 A^{SLR}_2 + (2\, I - \operatorname{ad}_{A^{SLR}_0})H^{SLR}_2 
 & = &\frac{1}{(4\pi)^2}\left(
\begin{array}{cc}
 2 a_{11}^{SLR} + c_1^{SLR} & \frac{1}{(4\pi)^2}\left(\frac{6}{N^4}-\frac{15}{2 N^2}+\frac{3}{2}\right) \\
 4 a_{21}^{SLR}+ c_2^{SLR}& 2 a_{22}^{SLR}+c_3^{SLR} \\
\end{array}
\right)  \nn \\
& = &\frac{1}{(4\pi)^2}\left(
\begin{array}{cc}
0  &  \frac{6}{N^4}-\frac{15}{2 N^2}+\frac{3}{2} \\
 0 & 0 \\
\end{array}
\right) 
\eea
with:
\bea
c_1^{SLR} & = & \frac{1}{(4\pi)^2}\left(-\frac{2}{3 N^4}+\frac{17}{2 N^3}-\frac{14}{3 N^2}+\frac{47}{3 N}\right) \nn \\
c_2^{SLR} & = & \frac{1}{(4\pi)^2}\left(-\frac{21}{8}-\frac{6}{N^2}\right) \nn \\
c_3^{SLR} & = & \frac{1}{(4\pi)^2}\left(\frac{14}{3}-\frac{2}{3 N^4}-\frac{7}{2 N^3}-\frac{4}{N^2}-\frac{61 N}{6}+\frac{41}{3 N}\right)
\eea
The solutions are:
\be
H^{VLR}_2   =  \frac{1}{(4\pi)^2}\left(
\begin{array}{cc}
 \frac{1}{3 N^4}-\frac{17}{4 N^3}+\frac{7}{3 N^2}-\frac{47}{6 N} & 0 \\
 \frac{3}{2 N^3}+\frac{21}{8 N} & \frac{1}{3 N^4}+\frac{7}{4 N^3}+\frac{2}{N^2}+\frac{61 N}{12}-\frac{41}{6 N}-\frac{7}{3} \\
\end{array}
\right)
\ee
and:
\be
H^{SLR}_2  =  \frac{1}{(4\pi)^2}\left(
\begin{array}{cc}
 \frac{1}{3 N^4}-\frac{17}{4 N^3}+\frac{7}{3 N^2}-\frac{47}{6 N} & 0 \\
 \frac{3}{2 N^2}+\frac{21}{8} & \frac{1}{3 N^4}+\frac{7}{4 N^3}+\frac{2}{N^2}+\frac{61 N}{12}-\frac{41}{6 N}-\frac{7}{3} \\
\end{array}
\right)
\ee
The corresponding gauge-transformed connections $A'(g)$ in the Poincar\`e-Dulac-Levelt normal form read:
\bea
-\left[\frac{\gamma^{VLR}(g)}{\beta(g)}\right]' &=& -S^{VLR}_2(g)\frac{\gamma^{VLR}(g)}{\beta(g)}\left(S^{VLR}_2(g)\right)^{-1} + \dfrac{\partial S^{VLR}_2 (g)}{\partial g} \left(S^{VLR}_2(g)\right)^{-1} \nn \\ 
& = & \frac{1}{g}\left(\frac{\gamma_{0D}^{VLR}}{\beta_0} + g^2 A^{'VLR}_2\right) \nn \\
& = & \frac{1}{g}\left[ \left(
\begin{array}{cc}
 \frac{2}{N^2} & 0 \\
 0 & \frac{2}{N^2}-2 \\
\end{array}
\right) + \frac{g^2}{(4\pi)^2} \left(
\begin{array}{cc}
 0 & \frac{6}{N^3}+\frac{3 N}{2}-\frac{15}{2 N} \\
 0 & 0 \\
\end{array}
\right)\right]
\eea
and:
\bea
-\left[\frac{\gamma^{SLR}(g)}{\beta(g)}\right]' &=& -S^{SLR}_2(g)\frac{\gamma^{SLR}(g)}{\beta(g)}\left(S^{SLR}_2(g)\right)^{-1} + \dfrac{\partial S^{SLR}_2 (g)}{\partial g} \left(S^{SLR}_2(g)\right)^{-1} \nn \\
& = & \frac{1}{g}\left(\frac{\gamma_{0D}^{SLR}}{\beta_0} + g^2 A^{'SLR}_2\right) \nn \\
& = & \frac{1}{g}\left[\left(
\begin{array}{cc}
 \frac{2}{N^2} & 0 \\
 0 & \frac{2}{N^2}-2 \\
\end{array}
\right) + \frac{g^2}{(4\pi)^2} \left(
\begin{array}{cc}
 0 & \frac{6}{N^4}-\frac{15}{2 N^2}+\frac{3}{2} \\
 0 & 0 \\
\end{array}
\right)\right] 
\eea
As a consequence, the corresponding $Z(x,\mu)$ can be read from eq. \eqref{zsol2} with $k=1$. 

\subsection{Dimension 8 operators in large-$N$ YM theory}

We demonstrate by explicit computation that both the case (I) and the resonant condition of the case (II) are realized in the large-$N$ YM theory, and that the unitarity constraint (section \ref{5.4}) in the free conformal limit is satisfied as well. \par
We consider the dimension-$8$ gauge-invariant Hermitian scalar operators in SU($N$) YM theory \cite{G1,G2}:
\bea
\label{dim8set}
\mathcal{O}_{B841} & = & \f{1}{N^4}F^a_{\mu\sigma}F^{a \, \mu\rho}F^{b \, \sigma\nu}F^b_{\rho\nu} \;\;\;  \;\;\; \mathcal{O}_{B842} =  \f{1}{N^4}F^a_{\mu\sigma}F^{b \, \mu\rho}F^{b \, \sigma \nu}F^a_{\rho\nu} \nn \\
\mathcal{O}_{B843} & = &  \f{1}{N^4}F^a_{\mu\sigma}F^a_{\nu\rho}F^{b \, \sigma\mu}F^{b \, \rho\nu} \;\;\;  \;\;\; \mathcal{O}_{B844} =  \f{1}{N^4}F^a_{\mu\sigma}F^b_{\nu\rho}F^{a \, \sigma\mu}F^{b \, \rho\nu} \nn \\
\mathcal{O}_{B845} & = &  \f{1}{N^4}d_4^{abcd}F^a_{\mu\sigma}F^{b \, \mu\sigma}F^c_{\nu\sigma}F^{d \, \nu\rho} \;\;\; \;\;\; \mathcal{O}_{B846} =  \f{1}{N^4}d_4^{abcd}F^a_{\mu\sigma}F^{c \, \mu\rho}F^{b \, \nu\sigma}F^d_{\nu\rho} \nn \\
\mathcal{O}_{B847} & = &  \f{1}{N^4}d_4^{acbd}F^a_{\mu\sigma}F^{b \, \mu\sigma}F^c_{\nu\rho}F^{d \, \nu\rho} \;\;\; \;\;\; \mathcal{O}_{B848} =  \f{1}{N^4}d_4^{abdc}F^a_{\mu\sigma}F^{c \, \mu\rho}F^{b \, \nu\sigma}F^d_{\nu\rho}
\eea
where $F_{\mu\nu}^a$ is:
\be
F_{\mu\nu}^a = \partial_\mu A^a_\nu - \partial_{\nu} A^a_{\mu} - g f^{abc}A^b_{\mu}A^c_\nu
\ee
with:
\be
d_4^{abcd} = d^{abe}d^{dce}
\ee   
where $f^{abc}$, $d^{abc}$ are:
\bea
&&\left[T^a, T^b\right] = i f^{abc} T^c \nn \\
&& \left\{T^a,T^b\right\} = \frac{1}{N}\delta^{ab} I + d^{abc}T^c
\eea
with the generators, $T^a$, of the Lie algebra of SU($N$) in the fundamental representation normalized as:
\bea
\Tr(T^aT^b)=\frac{1}{2} \delta^{ab}
\eea
We refer to the operators $\mathcal{O}_{B841}\cdots\mathcal{O}_{B844}$ and $\mathcal{O}_{B845}\cdots \mathcal{O}_{B848}$ as to double-trace and single-trace operators respectively. They mix among themselves under renormalization \cite{G1,G2}:
\be
\mathcal{O} = Z\mathcal{O}_B
\ee 
where $\mathcal{O}$ is the column vector of renormalized operators, whose transpose, $\mathcal{O}^T$, reads:
\be
\mathcal{O}^T = \left(\mathcal{O}_{841} \, \mathcal{O}_{8412} \, \mathcal{O}_{843} \,\mathcal{O}_{844} \,\mathcal{O}_{845} \,\mathcal{O}_{846} \,\mathcal{O}_{847} \,\mathcal{O}_{848}\right)
\ee
with $\mathcal{O}_B$ the vector of the bare operators.\par
The corresponding $\gamma_0$ reads \cite{G2}:
\bea \label{gammaDim8}
\gamma_0 = \sum_{k=0}^{3} \frac{1}{N^k} \gamma_{0k}
\eea
For the matrix of $2$-point correlators in the free conformal limit we get \cite{BB1}:
\bea \label{Gdim8}
G^{(2)}(x) = \frac{1}{(x{^2})^{8}}G_0
\eea
with:
\bea
G_0 = \sum_{k=0}^{6} \frac{1}{N^k} G_{0k}
\eea
We only report the leading-order terms, $\gamma_{00}$ and $G_{00}$, in the large-$N$ expansion \cite{BB1}:
\bea
\gamma_{00} & = & \dfrac{1}{(4\pi)^2} \left(
\begin{array}{cccccccc}
 0 & 0 & 0 & -\frac{11}{6} & 0 & 0 & 0 & 0 \\
 -\frac{14}{3} & -\frac{10}{3} & 4 & \frac{1}{6} & 0 & 0 & 0 & 0 \\
 -\frac{28}{3} & 8 & \frac{2}{3} & \frac{1}{3} & 0 & 0 & 0 & 0 \\
 0 & 0 & 0 & -\frac{22}{3} & 0 & 0 & 0 & 0 \\
 0 & 0 & 0 & 0 & -\frac{5}{2} & -6 & -\frac{2}{3} & \frac{16}{3} \\
 0 & 0 & 0 & 0 & -\frac{4}{3} & -1 & 1 & -1 \\
 0 & 0 & 0 & 0 & -\frac{25}{12} & -\frac{19}{3} & 1 & \frac{16}{3} \\
 0 & 0 & 0 & 0 & -\frac{5}{6} & -\frac{8}{3} & 2 & -\frac{4}{3} \\
\end{array}
\right) \nn \\
G_{00} & = & \dfrac{1}{\pi^8}\left(
\begin{array}{cccccccc}
 576 & 384 & 768 & 1152 & 0 & 0 & 0 & 0 \\
 384 & 768 & 1152 & 384 & 0 & 0 & 0 & 0 \\
 768 & 1152 & 2688 & 768 & 0 & 0 & 0 & 0 \\
 1152 & 384 & 768 & 4608 & 0 & 0 & 0 & 0 \\
 0 & 0 & 0 & 0 & 5376 & 1920 & 3456 & 1536 \\
 0 & 0 & 0 & 0 & 1920 & 1056 & 1728 & 960 \\
 0 & 0 & 0 & 0 & 3456 & 1728 & 4416 & 1920 \\
 0 & 0 & 0 & 0 & 1536 & 960 & 1920 & 1152 \\
\end{array}
\right)
\eea
Hence, to the leading large-$N$ order, single- and double-trace operators only mix separately among themselves \cite{BB1}.\par
According to eq. \eqref{2.35}, $\gamma_{00}$ and $G_{00}$ are simultaneously diagonalizable by the global gauge transformation \cite{BB1}:
\begin{small}
\be
S = \left(
\begin{array}{cccccccc}
 0 & -\frac{2}{3} & \frac{1}{3} & \frac{5 \sqrt{13}}{6}+3 & 0 & 0 & 0 & 0 \\
 0 & -\frac{2}{3} & \frac{1}{3} & 3-\frac{5 \sqrt{13}}{6} & 0 & 0 & 0 & 0 \\
 -2 & \frac{2}{3} & \frac{2}{3} & \frac{1}{3} & 0 & 0 & 0 & 0 \\
 2 & 0 & 0 & -\frac{1}{2} & 0 & 0 & 0 & 0 \\
 0 & 0 & 0 & 0 & \frac{1}{656} \left(41-3 \sqrt{41}\right) & \frac{7}{2 \sqrt{41}}-\frac{1}{2} & \frac{1}{8}-\frac{3}{8 \sqrt{41}} & \frac{1}{4}-\frac{11}{4 \sqrt{41}} \\
 0 & 0 & 0 & 0 & \frac{101}{16 \sqrt{697}}-\frac{1}{16} & \frac{1}{2}-\frac{13}{2 \sqrt{697}} & -\frac{1}{8}-\frac{75}{8 \sqrt{697}} & \frac{1}{4}+\frac{75}{4 \sqrt{697}} \\
 0 & 0 & 0 & 0 & \frac{1}{656} \left(3 \sqrt{41}+41\right) & -\frac{1}{2}-\frac{7}{2 \sqrt{41}} & \frac{1}{8}+\frac{3}{8 \sqrt{41}} & \frac{1}{4}+\frac{11}{4 \sqrt{41}} \\
 0 & 0 & 0 & 0 & -\frac{1}{16}-\frac{101}{16 \sqrt{697}} & \frac{1}{2}+\frac{13}{2 \sqrt{697}} & \frac{75}{8 \sqrt{697}}-\frac{1}{8} & \frac{1}{4}-\frac{75}{4 \sqrt{697}} \\
\end{array}
\right)
\ee
\end{small}
Therefore:
\begin{small}
\be
\label{gDiag}
\gamma'_{00} = \dfrac{1}{(4\pi)^2}\left(
\begin{array}{cccccccc}
 -\frac{22}{3} & 0 & 0 & 0 & 0 & 0 & 0 & 0 \\
 0 & -\frac{22}{3} & 0 & 0 & 0 & 0 & 0 & 0 \\
 0 & 0 & \frac{14}{3} & 0 & 0 & 0 & 0 & 0 \\
 0 & 0 & 0 & 0 & 0 & 0 & 0 & 0 \\
 0 & 0 & 0 & 0 & -\frac{1}{6} \left(3 \sqrt{41}+5\right) & 0 & 0 & 0 \\
 0 & 0 & 0 & 0 & 0 & -\frac{1}{12} \left(\sqrt{697}+13\right) & 0 & 0 \\
 0 & 0 & 0 & 0 & 0 & 0 & \frac{1}{6} \left(3 \sqrt{41}-5\right) & 0 \\
 0 & 0 & 0 & 0 & 0 & 0 & 0 & \frac{1}{12} \left(\sqrt{697}-13\right) \\
\end{array}
\right)  
\ee  
\end{small}
and:
\begin{scriptsize}
\be
\label{A00D}
G'_{00} = \dfrac{1}{\pi^8}\left(
\begin{array}{cccccccc}
 1280 \left(18 \sqrt{13}+65\right) & 0 & 0 & 0 & 0 & 0 & 0 & 0 \\
 0 & 1280 \left(65-18 \sqrt{13}\right) & 0 & 0 & 0 & 0 & 0 & 0 \\
 0 & 0 & 1280 & 0 & 0 & 0 & 0 & 0 \\
 0 & 0 & 0 & 1152 & 0 & 0 & 0 & 0 \\
 0 & 0 & 0 & 0 & 144-\frac{816}{\sqrt{41}} & 0 & 0 & 0 \\
 0 & 0 & 0 & 0 & 0 & 432+\frac{7056}{\sqrt{697}} & 0 & 0 \\
 0 & 0 & 0 & 0 & 0 & 0 & 144+\frac{816}{\sqrt{41}} & 0 \\
 0 & 0 & 0 & 0 & 0 & 0 & 0 & 432-\frac{7056}{\sqrt{697}} \\
\end{array}
\right)
\ee
\end{scriptsize}
Interestingly, the system above satisfies \cite{BB1} the resonant condition for some eigenvalues in the double-trace sector:
\bea
&&\frac{\gamma'_{004}}{\beta_0} - \frac{\gamma'_{001}}{\beta_0} = 2 \nn \\
&& \frac{\gamma'_{004}}{\beta_0} - \frac{\gamma'_{002}}{\beta_0} = 2
\eea
Moreover, despite $\gamma_{00}$ would be potentially nondiagonalizable because of the two coinciding eigenvalues in eq. \eqref{gDiag},
it is actually diagonalizable -- and $G_{00}$ as well -- according to the unitarity constraint (section \ref{5.4}) in the free conformal limit. 
Moreover, we verify that the eigenvalues of $G_{00}$ \cite{BB1} are all positive numbers:
\begin{small}
\bea
\label{A00E}
G'_{001} & = & \dfrac{1}{\pi^8}1280 \left(18 \sqrt{13}+65\right) = 17.52 \ldots \nn \\
G'_{002} & = & \dfrac{1}{\pi^8}1280 \left(65-18 \sqrt{13}\right) = 0.013 \ldots \nn \\
G'_{003} & = & \dfrac{1}{\pi^8}1280 = 0.1349 \ldots  \nn \\
G'_{004} & = & \dfrac{1}{\pi^8}1152 = 0.1214 \ldots  \nn \\
G'_{005}& = & \dfrac{1}{\pi^8}\left(144-\frac{816}{\sqrt{41}}\right) = 0.001 \ldots  \nn \\
G'_{006} & = & \dfrac{1}{\pi^8}\left( 432+\frac{7056}{\sqrt{697}}\right) = 0.017 \ldots  \nn \\
G'_{007}& = & \dfrac{1}{\pi^8}\left( 144+\frac{816}{\sqrt{41}}\right) = 0.028 \ldots  \nn \\
G'_{008} & = & \dfrac{1}{\pi^8}\left( 432-\frac{7056}{\sqrt{697}} \right) = 0.017 \ldots 
\eea
\end{small}
according to the aforementioned unitarity. 

\section{Acknowledgements}

The first named author also acknowledges the financial support from the European Union Horizon 2020 research and innovation programme: \emph{High precision multi-jet dynamics at the LHC} (grant agreement no. 772009).

\appendix

\section{Asymptotic versus exact correlators} \label{A}

We comment on the asymptotic versus exact form of the correlators in massless QCD-like theories.\par
The closed form of $Z(x, \mu)$ in eq. \eqref{Z} relies implicitly on the perturbative definition of $\gamma(g)$ and $\beta(g)$ that are believed to be formal series, 
at best asymptotic for $g \rightarrow 0$ thanks to the asymptotic freedom. \par
Correspondingly, the asymptotic solution of the Callan-Symanzik equation in eq. \eqref{2.200}, with $\mathcal{G}(x,g(\mu),\mu) \sim  \mathcal{G}(x,g(x))$, where $ \mathcal{G}(x,g(x))$ also relies on the RG-improvement of perturbation theory, is believed to be only asymptotic in the UV to the exact $2$-point correlator thanks to the asymptotic freedom.\par
The above statement may be verified directly in the large-$N$ limit of confining massless QCD-like theories following \cite{MBN}, where it has been shown how the aforementioned asymptotics works in the multiplicatively renormalizable case. \par
Indeed, as remarked in \cite{MB0}, nonperturbatively, according to the RG, massless QCD-like theories develop a nontrivial dimensionful scale that labels the RG trajectory -- the RG invariant -- $\Lambda_{RGI}$:
\bea \label{1}
 \Lambda_{RGI}  \sim   \mu \, e^{-\frac{1}{2\beta_0 g^2}} g^{-\frac{\beta_1}{ \beta_0^2}} c_0 (1+\sum_{n=1} c_n g^{2n})
 \eea
-- the only free parameter \cite{MBR,MBL} in the nonperturbative S matrix of confining massless QCD-like theories -- which any physical mass scale must be proportional to.\par
As a consequence, nonperturbatively in the large-$N$ limit of confining massless QCD-like theories \cite{H,V,Migdal,W}, the leading contribution to the exact Euclidean $2$-point correlators of gauge-invariant operators must be an infinite sum of free-field propagators \cite{Migdal,W}, with 
every mass in the propagators proportional to $ \Lambda_{RGI}$. \par
In the momentum representation, after the analytic continuation to Minkowski space-time, the sum of free propagators is a sum of pure poles, while the analytic continuation of the RG-improved \cite{MBN} perturbative solution of the Callan-Symanzik equation has only cuts \cite{MBN}, involving logs and loglogs \cite{MBN} of the momentum.\par
Therefore, the exact and all-order RG-improved $2$-point Euclidean correlators cannot coincide, otherwise their analytic continuations would coincide as well, though we have just shown that they do not. \par
Hence, RG-improved perturbation theory may only be UV asymptotic in large-$N$ confining QCD-like theories, and in fact, as remarked above, it is believed to be such because of the asymptotic freedom.

\end{document}